\documentclass[conference]{IEEEtran}
\IEEEoverridecommandlockouts

\usepackage[utf8]{inputenc} 
\usepackage[T1]{fontenc}    
\usepackage{hyperref}       
\usepackage{url}            
\usepackage{booktabs}       
\usepackage{amsfonts}       
\usepackage{nicefrac}       
\usepackage{microtype}      
\usepackage{xcolor}         
\usepackage{graphicx} 

\usepackage[linesnumbered,ruled,vlined,noend]{algorithm2e}
\usepackage[normalem]{ulem}
\usepackage{subcaption}
\usepackage{amsmath,amssymb,amsthm}
\usepackage{dirtytalk}
\usepackage{xspace}
\usepackage{tikz}
\usetikzlibrary{positioning, shapes.geometric, arrows.meta}
\usepackage{pifont}
\usepackage{wrapfig}
\usepackage{tcolorbox}

\usepackage{cite}
\usepackage{listings}
\lstdefinestyle{customprompt}{
  backgroundcolor=\color{gray!10},   
  basicstyle=\ttfamily\small,
  breaklines=true,                 
  captionpos=b,                    
  numbers=none,                    
  frame=single,                    
  rulecolor=\color{black},
  showstringspaces=false,
}
\usepackage{amsmath,amssymb,amsfonts}
\usepackage{textcomp}
\usepackage{cleveref}

\newcommand{\tool}{\texttt{BinaryShield}\xspace}

\usepackage{fancyhdr}
\usepackage{xcolor}

\makeatletter
\def\ps@IEEEtitlepagestyle{%
    \fancyhf{}
    \fancyhead[C]{\textcolor{blue}{\textbf{Accepted at the 2026 IEEE Conference on Secure and Trustworthy Machine Learning (SaTML)}}}
    \fancyfoot[C]{\thepage}
    \renewcommand{\headrulewidth}{0pt}
    \renewcommand{\footrulewidth}{0pt}
}
\makeatother

\def\BibTeX{{\rm B\kern-.05em{\sc i\kern-.025em b}\kern-.08em
T\kern-.1667em\lower.7ex\hbox{E}\kern-.125emX}}
\begin{document}

\title{BinaryShield: Cross-Service Threat Intelligence in LLM Services using Privacy-Preserving Fingerprints}

\author{\IEEEauthorblockN{Waris Gill}
  \IEEEauthorblockA{
    \textit{Microsoft}\\
    Redmond, USA \\
  warisgill@microsoft.com}
  \and
  \IEEEauthorblockN{Natalie Isak}
  \IEEEauthorblockA{
    \textit{Microsoft}\\
    New York, USA \\
    natalieisak@microsoft.com}
  \and
  \IEEEauthorblockN{Matthew Dressman}
  \IEEEauthorblockA{
    \textit{Microsoft}\\
    Redmond, USA \\
  matthew.dressman@microsoft.com}
  
}

\maketitle

\pagestyle{fancy}
\fancyhf{}
\fancyhead[C]{\textcolor{blue}{\textbf{Accepted at the 2026 IEEE Conference on Secure and Trustworthy Machine Learning (SaTML)}}}
\fancyfoot[C]{\thepage}
\renewcommand{\headrulewidth}{0pt}
\renewcommand{\footrulewidth}{0pt}

\begin{abstract}
    The widespread deployment of LLMs across enterprise services has created a critical security blind spot. Organizations operate multiple LLM services handling billions of queries daily, yet regulatory compliance boundaries prevent these services from sharing threat intelligence about prompt injection attacks, the top security risk for LLMs. When an attack is detected in one service, the same threat may persist undetected in others for months, as privacy regulations prohibit sharing user prompts across compliance boundaries.

    We present \tool, \emph{the first privacy-preserving threat intelligence system that enables secure sharing of attack fingerprints across compliance boundaries.}
    \tool transforms suspicious prompts through a unique pipeline combining PII redaction, semantic embedding, binary quantization, and randomized response mechanism to potentially generate privacy-preserving fingerprints that preserve attack patterns while providing privacy.
    Our evaluations demonstrate that \tool
    achieves an F1-score of 0.94, significantly outperforming SimHash (0.77), the privacy-preserving baseline, while achieving storage reduction and 38x faster similarity search compared to dense embeddings.

\end{abstract}

\section{Introduction}
\label{sec:introduction}

The rapid adoption of Large Language Models (LLMs) in digital services is driving a paradigm shift in human-computer interaction.
Companies now operate many LLM-based services across diverse domains, collectively handling billions of queries each day.
Typically, organizations like Microsoft maintain multiple, logically isolated (siloed) LLM services, such as enterprise AI assistants, consumer chat applications, API-based LLM services, and developer-focused coding agents.
These services are siloed because they are tailored to meet distinct business objectives and end-user requirements, necessitating separate operational boundaries.
Each service has distinct model stacks, logging pipelines, and compliance boundaries with strong internal data governance policies to protect user data and privacy. In addition to internal privacy regulations, these services are also subject to governmental regulations such as GDPR and HIPAA, which mandate strict data handling practices to protect user privacy and sensitive information.
While this siloed architecture is crucial for maintaining operational independence, it has an unintended consequence: fragmented security telemetry and a weakened collective defense.
With little correlation or data sharing between services, organizations are left with a disjointed security posture as each service operates independently, responding to threats in isolation. \emph{This fragmentation creates a significant challenge for incident response teams, as they lack a unified view of the threat landscape across their LLM services.}

\noindent{\textbf{Prompt Injection and Current Defenses.}} The new interaction model, based on natural language prompts, introduces a critical security risk: prompt injection attacks.
Often described as the \say{SQL injection} of AI, prompt injection is recognized as the top threat in the OWASP 2025 Top 10 for LLMs. Prompt injection exploits LLMs’ inability to distinguish legitimate user instructions from malicious commands embedded in input. These attacks can manipulate system prompts to leak private data, execute malicious code via tool interfaces, or generate harmful content. With emerging technologies like Model Context Protocol (MCP) and  autonomous agents, the impact of prompt injection extends beyond text manipulation to arbitrary code execution and potential system compromise~\cite{thehackernewsZeroClickVulnerability}.
Organizations and researchers have developed various defense mechanisms against these attacks~\cite{abdelnabi2025get, abdelnabi2025llmail}.
Furthermore, more than one defense is deployed to enhance security within a compliance boundary~\cite{abdelnabi2025llmail}.
However, a critical limitation stems directly from privacy regulations that prevent the sharing of user prompts between compliance boundaries (i.e., services). 

\noindent{\textbf{Motivation and Problem Statement.}}
The current defense paradigm is compartmentalized, with each service responding independently to detected attacks. When a new attack vector is identified (typically through media reports, user complaints, or post-breach analysis), security teams patch their systems (e.g., update prompts classifiers). However, this knowledge remains confined to the affected service. 
Furthermore, attacks are inevitable because existing defenses are probabilistic and can be bypassed~\cite{Debenedetti2025defeating, costa2025securingaiagentsinformationflow}.
Thus, organizations need retrospective systems to pinpoint the perpetrator and timing of incidents.

This reactive, siloed approach prevents organization-wide visibility into the threat landscape. An attack discovered in one service today may have existed undetected in another for months. Security analysts lack tools to search for similar historical attack patterns across service boundaries, creating a \say{correlation gap}. As a result, they cannot fully assess the scale and sophistication of an attack, identify its variants, or proactively protect all services from emerging threats.
Traditional malware defense addressed similar coordination challenges decades ago by exchanging signatures: antivirus engines share hash or pattern-based fingerprints of malicious binaries without revealing proprietary information~\cite{li2019reading}. To our knowledge, no comparable, privacy-preserving, and practically deployable threat intelligence mechanism exists for natural-language prompts in LLM services. This paper addresses the following problem: \emph{How can an organization securely share threat intelligence about prompt injection attacks across its compliance boundaries without violating privacy regulations?}

\noindent{\textbf{Ideal Characteristics.}}
A privacy-preserving fingerprinting mechanism for prompt injection attacks is essential for secure information sharing across compliance boundaries. Such a fingerprint must encode the semantic core of the malicious prompt, capturing its essential characteristics beyond surface tokens. 
It should be privacy preserving, so the original input (prompt) cannot be recovered. The technique must support secure approximate matching for \say{find-similar} searches across data stores and remain lightweight for real-time generation and distribution.
When an attack is detected in one service, the fingerprint can be broadcast to peer services. These peers can then search historical logs for related incidents, flag live traffic, and train local defenses, enabling proactive, collaborative security. Cross-boundary threat intelligence will enable faster patching without compromising user privacy.

\noindent{\textbf{Our Contributions.}} We propose the concept of \emph{threat intelligence} across compliance boundaries in LLM services. 
We materialize this idea in \tool, the first privacy-preserving fingerprinting technique for prompt injection attacks that enables secure information sharing between otherwise siloed services. 
When a service's defense mechanism detects a potential prompt-injection threat, \tool generates a privacy-preserving, lightweight fingerprint that captures the malicious prompt's semantic essence, allowing actionable threat signals to be shared without compromising privacy. \emph{Note}, we use \emph{prompt injection} as a broader term, similar to its use in the OWASP Top 10 for LLMs, where jailbreaking is treated as a form of prompt injection~\cite{LLM012025PromptInjection}.

Our approach balances the need for threat intelligence sharing with user privacy and regulatory compliance. \tool achieves this through a multi-stage process that progressively removes identifying information while retaining the essential semantic characteristics for effective threat detection.
The process begins with personally identifiable information (PII) redaction, where \tool removes sensitive data such as social security numbers, names, and other identifying markers from input prompts. This initial privacy layer is crucial for protecting users whose prompts may be incorrectly flagged as malicious by automated defenses. However, PII redaction alone is insufficient, as the remaining text may still contain contextual information that could compromise user privacy when shared externally.
To capture the semantic essence of potentially malicious prompts, \tool generates high-dimensional embeddings from the \emph{redacted text}. 
These embeddings, which represent text as dense floating-point vectors, achieve state-of-the-art performance in natural language processing tasks by encoding semantic relationships in vector space.
However, embeddings are not secure for sharing across compliance boundaries. 
Recent research demonstrates that embeddings pose significant privacy risks, as they can allow adversaries to reconstruct original input text~\cite{forbesSystemsVector, li2023sentence, tragoudaras2025information, owasp2025vectorEmbedding}. 
Moreover, traditional embeddings present practical challenges for large-scale threat intelligence system. 
Each dimension typically requires 32 bits of storage, and similarity computations using cosine similarity or dot products demand substantial computational resources, often necessitating GPU acceleration for efficient threat matching.

\tool addresses these fundamental challenges through two innovative non-reversible transformations by exploiting insights from quantization and differential privacy. 
First, \tool performs binary quantization on the floating-point embeddings, converting each dimension to a single bit based on its sign, assigning 1 for positive values and 0 for negative values. 
This quantization achieves remarkable efficiency gains, reducing storage requirements while simultaneously enhancing privacy. 
The key insight underlying this approach is that by discarding magnitude information and retaining only directional information, we make prompt reconstruction exponentially more difficult while preserving sufficient semantic structure for threat sharing and detection.
To render original prompt reconstruction practically impossible, \tool applies differential privacy to the binary embeddings.
Leveraging the insights from the principle of randomized response~\cite{warner1965randomized, rappor, Wang2017Locally}, \tool independently flips each bit of the binary vector with a precisely calibrated probability.
This noise addition fundamentally alters the binary vector representation while preserving its calibrated utility for threat correlation, ensuring that adversaries cannot reverse-engineer the original prompt from the shared fingerprint. 
The resulting binary vector is efficient in storage and computation, enabling rapid Hamming distance~\cite{manku2007detecting} similarity checks. 
More importantly, after these privacy-preserving transformations, the resulting binary vector becomes safe to share across compliance boundaries, enabling effective cross-service threat intelligence without compromising user privacy or regulatory compliance.
\begin{figure}[t]
    \centering
    \includegraphics[width=0.95\linewidth]{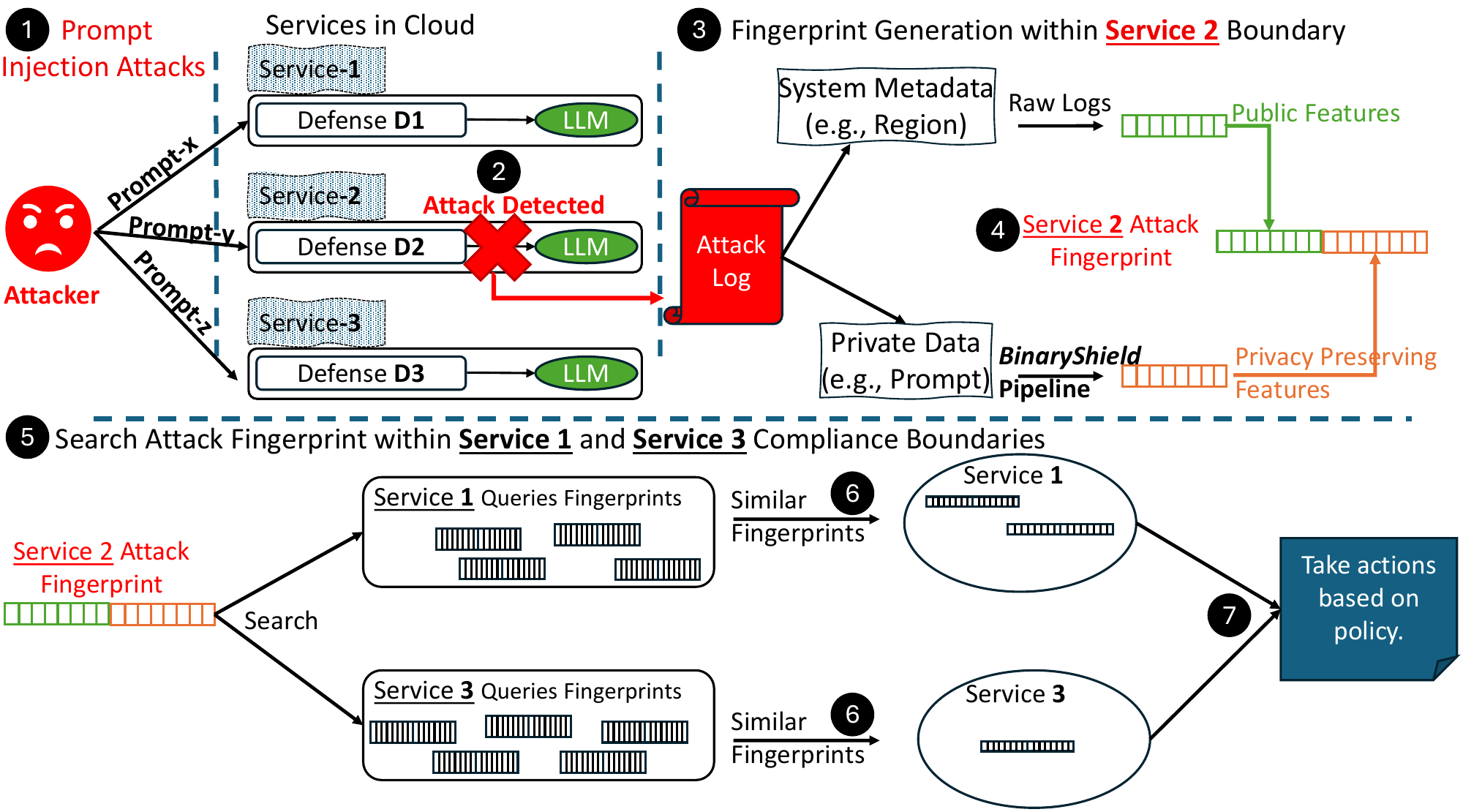}
    \caption{\tool system design. Suspicious prompts are processed within the compliance boundary to generate privacy-preserving fingerprints, which are then shared across services for collaborative threat detection.}
    \label{fig:system-architecture}
    \vspace{-4mm}  
\end{figure}

\noindent\textbf{Evaluations.} We conduct comprehensive evaluation of \tool spanning threat correlation effectiveness, privacy calibration, scalability, and operational efficiency (Section~\ref{sec:evaluation}). 
\tool significantly outperforms privacy-preserving baseline across attack variants (17-point F1 advantage on sophisticated attacks) and exhibits smooth privacy-utility trade-offs with theoretical noise alignment. 
In real-world enterprise-scale scenarios, \tool achieves 79.2\% threat correlation accuracy (93\% of non-private baseline performance) alongside efficiency gains (38x search speedup).

\section{\tool Design}
\label{sec:design}
\begin{figure*}[t]    
\centering  
\resizebox{\linewidth}{!}{%
\begin{tikzpicture}[    
    node distance=2.5cm,    
    box/.style={rectangle, draw, thick, minimum height=1cm, minimum width=3cm, align=center, rounded corners=3pt},    
    privacy/.style={box, fill=blue!15},    
    transform/.style={box, fill=green!15},    
    output/.style={box, fill=orange!15},    
    arrow/.style={->, thick, >=stealth},    
    label/.style={font=\footnotesize, align=center}    
]    
    
\node[box] (input) {Potential Prompt\\Injection};    
    
\node[privacy, right=of input] (pii) {PII Redaction\\Module};    
    
\node[transform, right=of pii] (embed) {Semantic\\Embedding\\Generation};    
    
\node[transform, right=of embed] (quant) {Binary\\Quantization};    
    
\node[privacy, right=of quant] (dp) {Randomized \\Response\\(Bit Flipping)};    
    
\node[output, right=of dp] (fingerprint) {Privacy-Preserving\\Binary Fingerprint};    
    
\draw[arrow] (input) -- node[above, label] {Raw prompt} (pii);    
\draw[arrow] (pii) -- node[above, label] {Redacted text} (embed);    
\draw[arrow] (embed) -- node[above, label] {Float vector\\$\mathbf{e} \in \mathbb{R}^d$} (quant);    
\draw[arrow] (quant) -- node[above, label] {Binary vector\\$\mathbf{b} \in \{0,1\}^d$} (dp);    
\draw[arrow] (dp) -- node[above, label] {Noisy binary\\$\tilde{\mathbf{b}} \in \{0,1\}^d$} (fingerprint);    
       
\node[align=center, font=\footnotesize] at ([yshift=1.0cm]quant.north) {    
    \textit{Transformation:}\\[0.2em]    
    $b_i = \begin{cases}    
    1 & \text{if } e_i > 0\\    
    0 & \text{otherwise}    
    \end{cases}$    
};    
    
\node[align=center, font=\footnotesize] at ([yshift=1.0cm]dp.north) {    
    \textit{Privacy mechanism:}\\[0.2em]    
    $\Pr[\tilde{b}_i = b_i] = \frac{e^\alpha}{e^\alpha + 1}$    
};    
    
\end{tikzpicture}
}
\caption{\tool approach for privacy-preserving fingerprint generation for cross-service threat intelligence sharing. The pipeline transforms potential prompt injections through PII redaction, semantic embedding, binary quantization, and differential privacy to produce shareable fingerprints that preserve privacy while enabling threat correlation across compliance boundaries.}    
\label{fig:BinaryShield-Privacy-Pipeline}
\vspace{-4mm}      
\end{figure*}
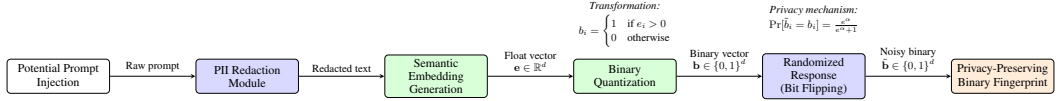  
\begin{algorithm}[t]
{\footnotesize
    \SetKwInput{KwIn}{Input}
    \SetKwInput{KwOut}{Output}
    \SetKwFunction{PIIRedact}{PII\_Redact}
    \SetKwFunction{Embed}{Embed}
    \SetKwFunction{Quantize}{Quantize}
    \SetKwFunction{RandResp}{RandomizedResponse}
    \SetKwFunction{Broadcast}{Broadcast}
    \SetKwFunction{Correlate}{Correlate}

    \KwIn{Suspicious prompt $y$; system metadata $m$; privacy parameter $\alpha$;  set of peer services $\mathcal{S}$}
    \KwOut{Privacy-preserving fingerprint $f$ broadcast to $\mathcal{S}$}

    $y_{redacted} \leftarrow$ \PIIRedact{$y$} \tcp*[r]{Remove all PII, replace with placeholders}

    $\mathbf{e} \leftarrow$ \Embed{$y_{redacted}$} \tcp*[r]{Compute semantic embedding vector}

    $\mathbf{b} \leftarrow []$ \tcp*[r]{Initialize empty binary vector}
    \For{$i \leftarrow 1$ \KwTo $d$}{
        $b_i \leftarrow \begin{cases}
                1 & \text{if } e_i > 0 \\
                0 & \text{otherwise}
            \end{cases}$ \;
        Append $b_i$ to $\mathbf{b}$ \;
    }

    $p_\text{keep} \leftarrow \frac{e^{\alpha}}{e^{\alpha} + 1}$ \tcp*[r]{Probability of keeping the bit}
    $\tilde{\mathbf{b}} \leftarrow []$ \tcp*[r]{Initialize empty privatized vector}
    \For{$i \leftarrow 1$ \KwTo $d$}{
        Sample $u_i \sim \mathrm{Uniform}(0,1)$ \;
        \eIf{$u_i < p_\text{keep}$}{
            $\tilde{b}_i \leftarrow b_i$ \tcp*[r]{Keep the true bit}
        }{
            $\tilde{b}_i \leftarrow 1 - b_i$ \tcp*[r]{Flip the bit}
        }
        Append $\tilde{b}_i$ to $\tilde{\mathbf{b}}$ \;
    }

    $f \leftarrow$ Concatenate$(\tilde{\mathbf{b}}, m)$ \tcp*[r]{Combine privatized bits with non-private metadata}

    \Broadcast{$f$, $\mathcal{S}$} \tcp*[r]{Send fingerprint to peers}
    }
    \caption{\tool}
    \label{alg:binaryshield-pipeline}
\end{algorithm}

The proliferation of LLM-based services within organizations creates a fundamental tension between security effectiveness and privacy compliance.
While traditional malware defense systems successfully share threat signatures across boundaries~\cite{li2019reading}, no comparable mechanism exists for LLM services.
\tool addresses this gap by enabling organizations to share actionable threat intelligence about prompt injection attacks without violating privacy regulations. 

The system is designed to achieve four primary objectives: 
(1) \emph{Semantic Preservation}: capture the essential characteristics of malicious prompts that enable cross-service threat correlation; 
(2) \emph{Privacy Guarantee}: ensure that original prompts cannot be reconstructed from shared fingerprints, hence upholding regulations; 
(3) \emph{Computational Efficiency}: enable efficient fingerprint generation and search; and (4) \emph{Operational Compliance}: maintain strict separation of user data across compliance boundaries while enabling collaborative defense.
To this end, \tool transforms potentially malicious prompts through a series of privacy-preserving operations, ultimately producing compact binary fingerprints that can be safely shared between services.
The key insight underlying our approach is that prompt injection attacks, despite surface-level variations, share semantic patterns that can be captured and compared without revealing the original content.

\noindent{\textbf{High-Level Architecture.}} As a motivating example, consider a company that offers both \emph{Enterprise AI} and \emph{Consumer AI} services operating under separate compliance boundaries.
When \emph{Enterprise AI} detects a prompt injection attack, it cannot warn \emph{Consumer AI} due to regulations prohibiting raw prompt sharing, leaving \emph{Consumer AI} vulnerable to the same attack. 
Figure~\ref{fig:system-architecture} illustrates how \tool addresses this challenge.
An attacker issues semantically related prompts ($x,y,z$) to the services of an enterprise organization that reside in separate compliance boundaries (Figure~\ref{fig:system-architecture}-\ding{202}).
Each service applies their own defense mechanism ($D1$, $D2$, and $D3$) to incoming prompts.
Service-2 flags a prompt-injection attempt (Figure~\ref{fig:system-architecture}-\ding{203}).
After this step, \tool's fingerprinting pipeline is invoked in the service's compliance boundary (Figure~\ref{fig:system-architecture}-\ding{204}).
At this stage, the attack log is separated into two components.
The first and most critical component is private information: this contains the sensitive content of the prompt itself, holding the essential details of the attack.
Although critical for accurate event correlation, this data must remain within the service’s compliance boundary to safeguard privacy.
The suspicious prompt then undergoes a series of privacy-preserving transformations to generate a fingerprint within the attacked service's compliance boundary (Figure~\ref{fig:system-architecture}-\ding{204}).
The fingerprint generation process is detailed in Section~\ref{sec:one-way-fingerprint-generation}.
The second part consists of non-private system metadata, such as the tools used during response generation and the geographical region.
Since this metadata contains no sensitive data, it can be safely shared among peer services as part of complete signature.
Sharing these features enhances attack correlation by providing complementary evidence to surface relevant patterns pointing to attack source and channel, which helps reduce false positives and prioritize alerts.
For instance, matching system metadata across services, such as both accessing an email tool, significantly increases the confidence in the correlation of potential attacks.
Concatenating both parts forms a composite \emph{attack fingerprint} that discloses no readable text (Figure~\ref{fig:system-architecture}-\ding{205})
This fingerprint is then securely broadcast to peer services for threat correlation (Figure~\ref{fig:system-architecture}-\ding{206}).
Note that peer services (i.e., Services 1 and 3) use the exact same technique to fingerprint their local private queries and system metadata as used in Step \ding{204} of Figure~\ref{fig:system-architecture}.
For instance, after searching their logs, Service 1 and Service 3 independently find two and one similar attacks, respectively (Figure~\ref{fig:system-architecture}-\ding{207}).
Policy-driven response actions are then executed for each service based on the correlation results (Figure~\ref{fig:system-architecture}-\ding{208}).
\emph{Note that there is currently no public dataset available for system metadata.
    Therefore, our evaluation of \tool focuses exclusively on private information (prompt) fingerprinting.
    Nonetheless, including system metadata features in the overall design demonstrates the full architectural intent of \tool and highlights additional layers of security and context.}

\subsection{\tool Fingerprint Generation Design}
\label{sec:one-way-fingerprint-generation}
\tool design is guided by several key requirements. A primary goal is to ensure data privacy by transforming original prompts into fingerprints through a one-way process, making the reconstruction of the original prompt computationally difficult. Simultaneously, these fingerprints must preserve enough semantic information to be useful for identifying related attack patterns via approximate matching. Operationally, the system is built for computational efficiency to handle millions of daily queries with low latency and is designed with cross-boundary compatibility in mind, allowing fingerprints to be shared across services with varying internal regulatory requirements.
Algorithm~\ref{alg:binaryshield-pipeline} and Figure~\ref{fig:BinaryShield-Privacy-Pipeline} depicts the \tool's pipeline for generating privacy-preserving fingerprints of prompt injection attacks.
Below, we detail each component of the pipeline.

\subsubsection{PII Redaction}
\label{sec:pii-redaction}
\tool's pipeline initial stage directly targets privacy risks by identifying and removing personally identifiable information (PII) that may be present in flagged prompts.
While automated defenses aim to identify malicious content, they may inadvertently flag benign prompts containing sensitive user data.
\begin{figure}

    \centering
    \includegraphics[width=0.8\linewidth]{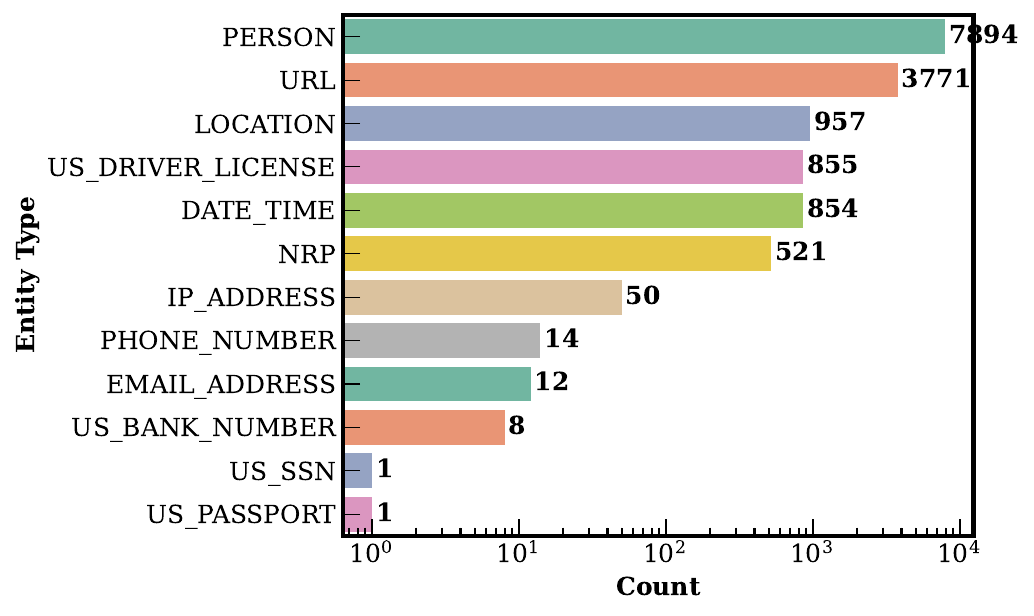}
    \caption{Distribution of PII entities in prompts flagged by \tool. Most common: person names.}
    \label{fig:pii-entity-counts}
    \vspace{-4mm}  
\end{figure}

The main challenge is removing PII while preserving the semantic structure of potential attacks. 
Our approach addresses this by detecting and redacting structured data (using~\cite{HomeMicr83:online}) such as social security numbers, credit card numbers, phone numbers, email addresses, person names, organizations, locations, and other identifying entities.
Identified PII is replaced with generic placeholders that maintain command structure, entity relationships, and linguistic patterns indicative of injection attempts.
For example, \say{Transfer \$5000 from John Smith's account 123456789} becomes \say{Transfer [AMOUNT] from [PERSON]'s account [ACCOUNT]}.
This balance ensures the redacted text retains sufficient information for downstream processing while protecting user privacy from the earliest stage of the pipeline.
This one-way redaction minimizes PII exposure before any downstream analysis. Figure~\ref{fig:pii-entity-counts} reports the PII entities and their respective counts that \tool flags in the prompt injection dataset from \cite{shen2024doanything}, used in \tool's evaluation (Section~\ref{sec:evaluation}). However, because redaction alone is insufficient to protect all sensitive information, the pipeline's next stage generates a privacy-preserving semantic fingerprint.

\begin{figure*}
    \centering
    \includegraphics[width=0.8\linewidth]{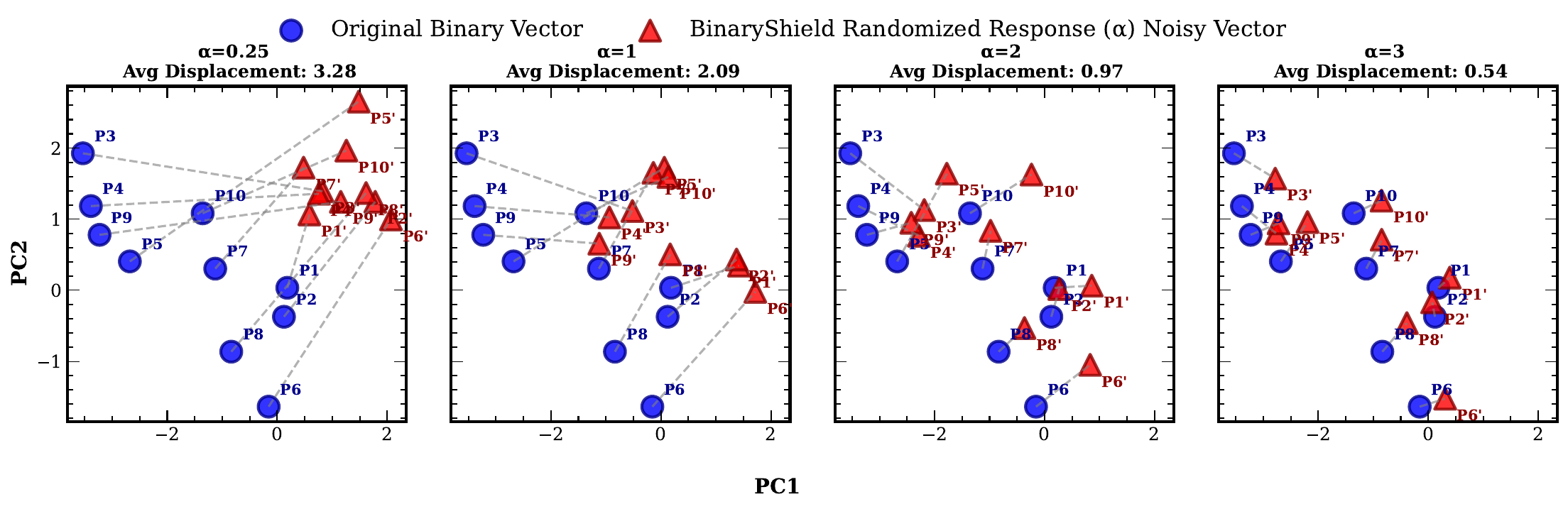}

    \caption{Impact of \tool's privacy parameter ($\alpha$). As $\alpha$ increases, privatized vector (red triangles) move closer to original vector, showing improved utility while maintaining differential privacy through controlled bit-flipping (i.e., $1-p$) noise.
    }
    \label{fig:visualize-binary-embeddings-pca}
    \vspace{-4mm}  
\end{figure*}

\subsubsection{Semantic Feature Extraction}
Following PII redaction, \tool must capture the semantic essence of the potentially malicious prompt in a format suitable for comparison and analysis. To this end, \tool employs state-of-the-art embedding models~\cite{warner-etal-2025-smarter, gill2025advancingsemanticcachingllms} to generate high-dimensional embeddings that encode the semantic relationships within the text. This transformation serves as the foundation for our semantic preservation requirement. 
This captures the meaning and intent of the potentially malicious prompt.
The embedding process transforms the redacted text into a dense vector $\mathbf{e} \in \mathbb{R}^d$, where $d$ is the embedding dimension.
The embedding space exhibits several properties that are crucial for threat detection.
First, similar attack patterns tend to cluster together in the embedding space, regardless of surface-level variations.
Second, the embeddings are robust to paraphrasing, so minor rewording or synonym substitution results in nearby embeddings, enabling the detection of attack variants.

\noindent{\textbf{Binary Quantization.}}
While semantic embeddings effectively capture prompt meaning, they pose significant challenges for privacy-preserving threat intelligence.
Dense embeddings can leak information about the original text~\cite{forbesSystemsVector, li2023sentence, tragoudaras2025information, owasp2025vectorEmbedding}.
They also require substantial computational resources for large-scale similarity searches.
Our binary quantization addresses both concerns.
We transform the continuous embedding $\mathbf{e} \in \mathbb{R}^d$ into a binary vector $\mathbf{b} \in \{0,1\}^d$ using sign-based quantization:

$$
    b_i = \begin{cases}
        1 & \text{if } e_i > 0 \\
        0 & \text{otherwise}
    \end{cases},
$$

where $e_i$ is the $i$-th dimension of the embedding vector and $b_i$ is the corresponding bit in the binary vector.
This transformation provides several critical benefits by design. 
First, it acts as a lossy compression mechanism. 
By mapping each continuous value in $\mathbb{R}$ to a binary state in $\{0,1\}$  based on its sign, the process discards all magnitude information. 
This many-to-one mapping makes it computationally difficult to infer the original floating-point vectors, thus enhancing data protection. 
Second, despite this information reduction, the high dimensionality ($d$) of the binary embedding preserves essential geometric properties needed for similarity detection. 
Finally, this method is highly storage-efficient, reducing the memory footprint for each dimension from 32 bits (for a float32) to just 1 bit, a 32x reduction.

\subsubsection{\tool's Randomized-Response Layer}
To build upon the baseline protections of data anonymization and binary quantization, our pipeline incorporates Local Differential Privacy (LDP) for a more formal privacy guarantee. 
By applying a randomized response mechanism~\cite{warner1965randomized,rappor,Wang2017Locally} to each bit before a fingerprint crosses the compliance boundary, we gain several critical advantages. 
This approach provides a strong, mathematically-defined upper bound on privacy loss for each fingerprint, ensuring plausible deniability with a formal guarantee of privacy regardless of any auxiliary information a person might possess. 
The guarantee is enforced locally before data leaves its origin, which is ideal for cross-boundary threat intelligence as it removes the need for a trusted central aggregator. Furthermore, unlike in the global DP, the magnitude of the required noise is independent of the dataset size, ensuring our privacy protection remains constant and does not degrade as the system scales to millions of daily fingerprints.

Consider a $d$-bit fingerprint $\mathbf{b}=(b_1,\dots,b_d)\!\in\!\{0,1\}^d$ and let $\alpha\!>\!0$ denote the per-bit privacy budget.
Classical randomized response publishes a perturbed vector $\tilde{\mathbf{b}}$ obtained by flipping each bit independently with probability $1-p$, where
\[
    p = \frac{e^{\alpha}}{e^{\alpha}+1} .
\]

The mechanism releases the randomized bit.
\[
    \tilde{b}_i \;=\;
    \begin{cases}
        b_i   & \text{with probability } p,    \\[6pt]
        1-b_i & \text{with probability } 1-p .
    \end{cases}
\]

This DP mechanism inevitably perturbs the fingerprint. The expected self-distance introduced by it is
\[
    \mathbb{E}\bigl[\mathrm{H}(\mathbf{b},\mathbf{\tilde{b}})\bigr] \;=\; (1-p)d,
\]
where $\mathrm{H}$ denotes Hamming distance and $d$ is the fingerprint dimensionality. The \emph{Hamming distance} between two binary vectors $\mathbf{b}, \mathbf{\tilde{b}} \in \{0,1\}^d$ is defined as the number of positions at which the corresponding bits differ:
\[
    \mathrm{H}(\mathbf{b}, \mathbf{\tilde{b}}) = \sum_{i=1}^d \mathbb{I}[b_i \ne  \tilde{b_i}] ,
\]
where $\mathbb{I}[\cdot]$ is the indicator function, which equals 1 if its argument is true and 0 otherwise.
Because $p$ remains close to one even for moderate $\alpha$, the added noise is sparse and, as we show experimentally in Section~\ref{sec:evaluation}, preserves enough structure for accurate approximate matching while providing the privacy guarantees demanded by cross-service threat intelligence.

The randomization budget $\alpha$ serves as a control parameter governing the privacy-utility tradeoff in \tool. 
When $\alpha$ approaches zero, the mechanism maximizes privacy protection (i.e., infinite privacy) but renders fingerprints unusable for correlation. 
As $\alpha$ increases, utility improves while maintaining privacy guarantees in a controlled manner.
Figure~\ref{fig:visualize-binary-embeddings-pca} illustrates this relationship through PCA visualization of binary vectors before and after applying the Randomized Response mechanism~\cite{warner1965randomized}. 
At $\alpha = 0.25$, privatized vectors (red triangles) appear completely random. 
As $\alpha$ increases (reducing noise), these vectors progressively align closer to their original non-private counterparts (blue circles). 
Notably, even at $\alpha = 3$, a non-zero average displacement between original and privatized vectors persists, making reconstruction of the original prompt challenging. 
This visualization confirms that larger $\alpha$ values preserve more of the vector's structure while maintaining formal differential privacy guarantees through controlled bit-flipping. 
In practice, higher privacy budgets may be appropriate for intra-service sharing where additional security measures already exist within organizational boundaries.

\subsubsection{\tool's Cross-Service Threat Correlation}
When a service detects a potential prompt injection, it generates a fingerprint using the \tool's pipeline.
For clarity, we denote the locally randomized-response fingerprint $\tilde{\mathbf b}\in\{0,1\}^d$ as $f$ when broadcasting and correlating across services (i.e., $f\approx \tilde{\mathbf b}$).
This fingerprint is broadcast asynchronously to participating services.
Upon receiving a fingerprint, services scan their historical logs for similar instances.
If matches are found, alerts are triggered and defenses can be updated proactively.
To maintain privacy and compliance, only aggregate match statistics are shared, and no specific prompt content is revealed.
This workflow enables rapid propagation of threat intelligence while respecting organizational boundaries.
Formally, let $\mathcal{S} = \{S_1, S_2, \ldots, S_N\}$ denote the set of $N$ services, each with its own compliance boundary.
Each service $S_i$ maintains a log of previously observed fingerprints $\mathcal{F}_i = \{f_i^{(1)}, f_i^{(2)}, \ldots, f_i^{(M_i)}\}$, where $f_i^{(m)} \in \{0,1\}^d$ is a $d$-dimensional binary fingerprint.
Suppose service $S_q$ detects a suspicious prompt and generates a fingerprint $f_q \in \{0,1\}^d$ using the \tool pipeline.
This fingerprint is securely broadcast to all other services $S_j$ ($j \neq q$).
Each recipient service $S_j$ performs the following search:

\[
    \mathcal{M}_j(f_q) = \left\{ f_j^{(m)} \in \mathcal{F}_j \;\middle|\; \mathrm{H}(f_q, f_j^{(m)}) \leq \tau \right\} ,
\]

where $\mathcal{M}_j(f_q)$ is the set of fingerprints belonging to service $S_j$ (i.e., fingerprints in $\mathcal{F}_j$) that match $f_q$ within a Hamming distance threshold $\tau$, and $\mathrm{H}(\cdot, \cdot)$ is the Hamming distance as previously defined.
This enables rapid, privacy-preserving propagation of threat intelligence across all $N$ services.
This protocol generalizes to any number of services.
For example, if Service 2 ($S_2$) generates a fingerprint $f_2$, it broadcasts $f_2$ to Services 1 and 3 ($S_1$, $S_3$), which independently search their logs for similar fingerprints.
If $S_1$ finds two matches and $S_3$ finds one, these services take action based on their policy and only the counts are reported back, enabling rapid, privacy-preserving propagation of threat intelligence across organizational boundaries.

\noindent{\textbf{Summary.}} \tool offers a scalable and privacy-preserving solution for sharing prompt injection threat intelligence across organizational compliance boundaries.
By combining PII redaction, semantic embedding, binary quantization, and differential privacy, the system achieves strong privacy protections without sacrificing detection performance or scalability.
The modular architecture allows for privacy parameters to be tuned to specific requirements, while efficient processing supports production-scale deployments.
As LLM adoption expands, \tool establishes a foundation for collaborative defenses against evolving attacks.

\section{Evaluation}
\label{sec:evaluation}

We present a comprehensive evaluation of \tool, structured around the following research questions:

\begin{itemize}
    \item \textbf{RQ1 (Detection Effectiveness).}
          How accurately does \tool match prompt injection attacks across systematically generated adversarial variants, from simple word substitutions to comprehensive semantic paraphrasing, compared to the SimHash~\cite{manku2007detecting} baseline?

    \item \textbf{RQ2 (Privacy-Utility Trade-off).} What is the quantitative relationship between \tool's randomized response mechanism ($\alpha$) and threat detection utility?

    \item \textbf{RQ3 (Noise Calibration \& Predictability).} Does the empirical self-Hamming distortion introduced by randomized response match the theoretical $(1-p)d$ curve across $\alpha$, letting us confidently choose a privacy setting?

    \item \textbf{RQ4 (Scalability).} Does \tool maintain consistent detection accuracy as corpus size scales from thousands to hundreds of thousands of entries, reflecting realistic enterprise deployment scenarios where malicious prompts are sparse within benign prompts?

    \item \textbf{RQ5 (Computational Efficiency).} What are the computational gains of \tool's fingerprints compared to dense embeddings, and do these improvements enable practical real-time threat correlation at enterprise scale?

\end{itemize}

\subsection{Evaluation Settings}

\subsubsection{Datasets}
We evaluate \tool using the recent prompt injection dataset from \cite{shen2024doanything}, which contains real-world attack and benign prompts. From this, we construct a comprehensive evaluation set simulating adversarial scenarios, enabling controlled and realistic assessment of threat intelligence. 

Given the inherent challenge of obtaining large-scale labeled datasets of semantically similar prompt injection variants, we develop a systematic methodology for generating controlled attack modifications to validate threat intelligence.
Our synthetic data generation process creates two primary types of variants: \emph{Word-Flipping Variants:} For each adversarial prompt, we randomly select $x$ words (where $x \in \{1, 3, 5, 10, 20\}$) longer than four characters and replace them with semantically equivalent synonyms using GPT-4o.
This simulates subtle adversarial modifications that maintain attack intent while altering surface-level tokens.
The language model is guided using the structured prompt listed in Appendix~\Cref{prompt:words-flipped}.
\emph{Paraphrase Variants:} We generate comprehensive paraphrased versions of prompts using GPT-4o while preserving the underlying attack intent. 
Appendix~\Cref{prompt:paraphrase} contains the prompt.
\emph{Benign Pairs:} Benign prompts are systematically paired such that each pair consists of two semantically unrelated prompts, ensuring they serve as appropriate negative examples for similarity detection algorithms (e.g., SimHash and \tool).
The resulting dataset undergoes rigorous filtering to ensure quality and consistency (e.g., remove duplicates, empty prompts).
The final dataset is balanced with equal numbers of attack and benign pairs, with attack pairs labeled as 1 and benign pairs labeled as 0.
This balanced approach ensures unbiased evaluation across both positive and negative cases.
Our synthetic generation methodology enables systematic evaluation across varying levels of attack sophistication, from minimal single-word modifications to comprehensive paraphrasing.

\subsubsection{Baseline}

Our evaluation compares \tool against SimHash~\cite{manku2007detecting}, a widely-adopted locality-sensitive hashing baseline.
\tool operates by first extracting semantic features using an embedding model. 
To demonstrate its flexibility with both open-source and proprietary options, we evaluate \tool using ModernBert~\cite{gill2025advancingsemanticcachingllms} and an OpenAI model.

\subsubsection{Evaluation Metrics}
We employ a comprehensive set of metrics to assess both detection performance and privacy-utility tradeoffs. For detection performance, we use precision, recall, accuracy, and F1-score across varying similarity thresholds to evaluate attack detection accuracy. Threshold analysis is performed using precision-recall curves to identify the optimal threshold for Hamming distance and to understand how performance is affected by threshold selection, as discussed in~\cite{manku2007detecting}. We also include confusion matrices for detailed analysis of true positives (TP), false positives (FP), true negatives (TN), and false negatives (FN) at optimal thresholds.

\begin{figure}[t]
    \begin{subfigure}[b]{0.45\textwidth}
        \includegraphics[width=\textwidth]{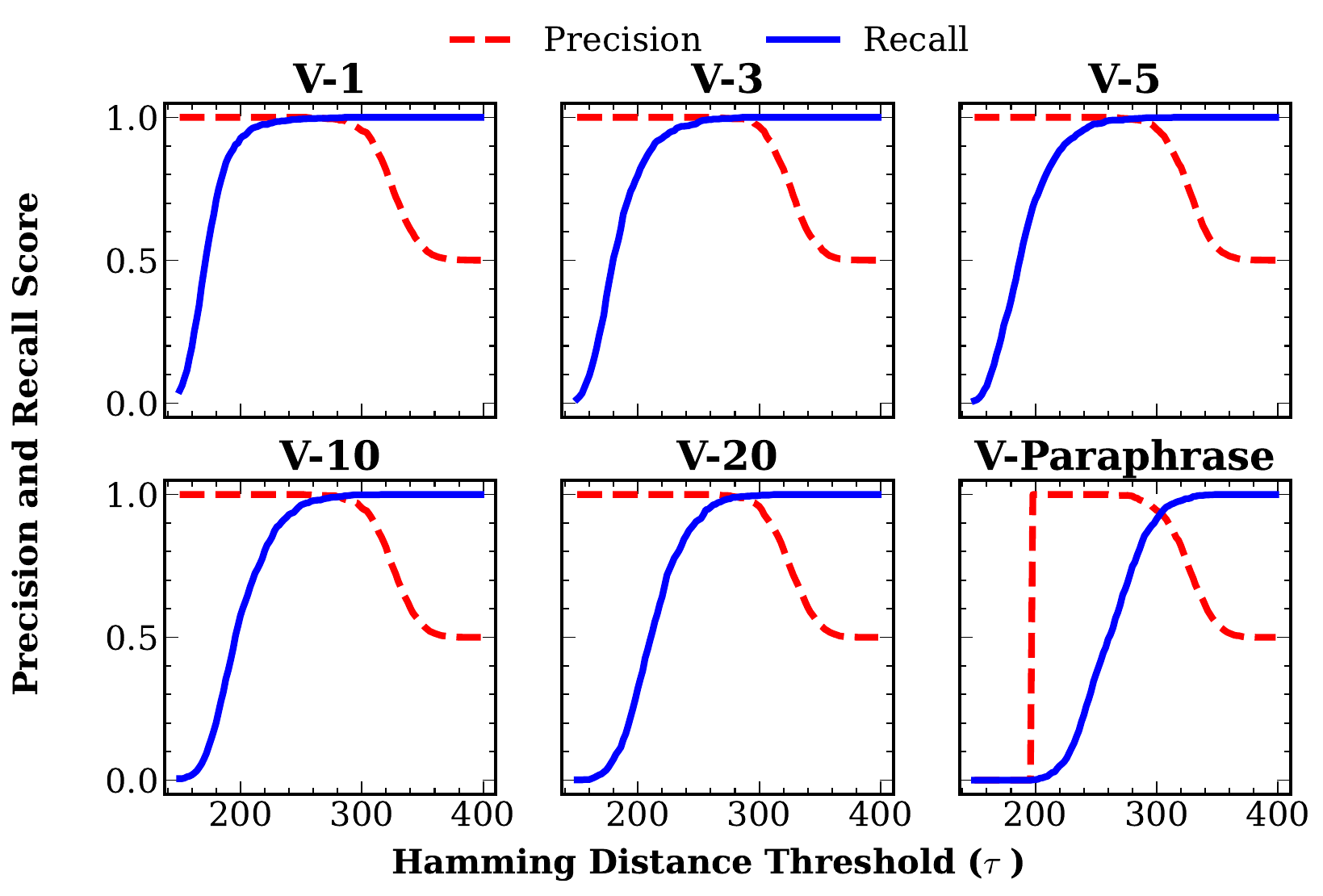}
        \caption{Precision-recall curves across attack variants.}
        \label{fig:binaryshield_pr}
    \end{subfigure}
    \hfill 
    \begin{subfigure}[b]{0.45\textwidth}
        \centering
            \centering
            \includegraphics[width=\linewidth]{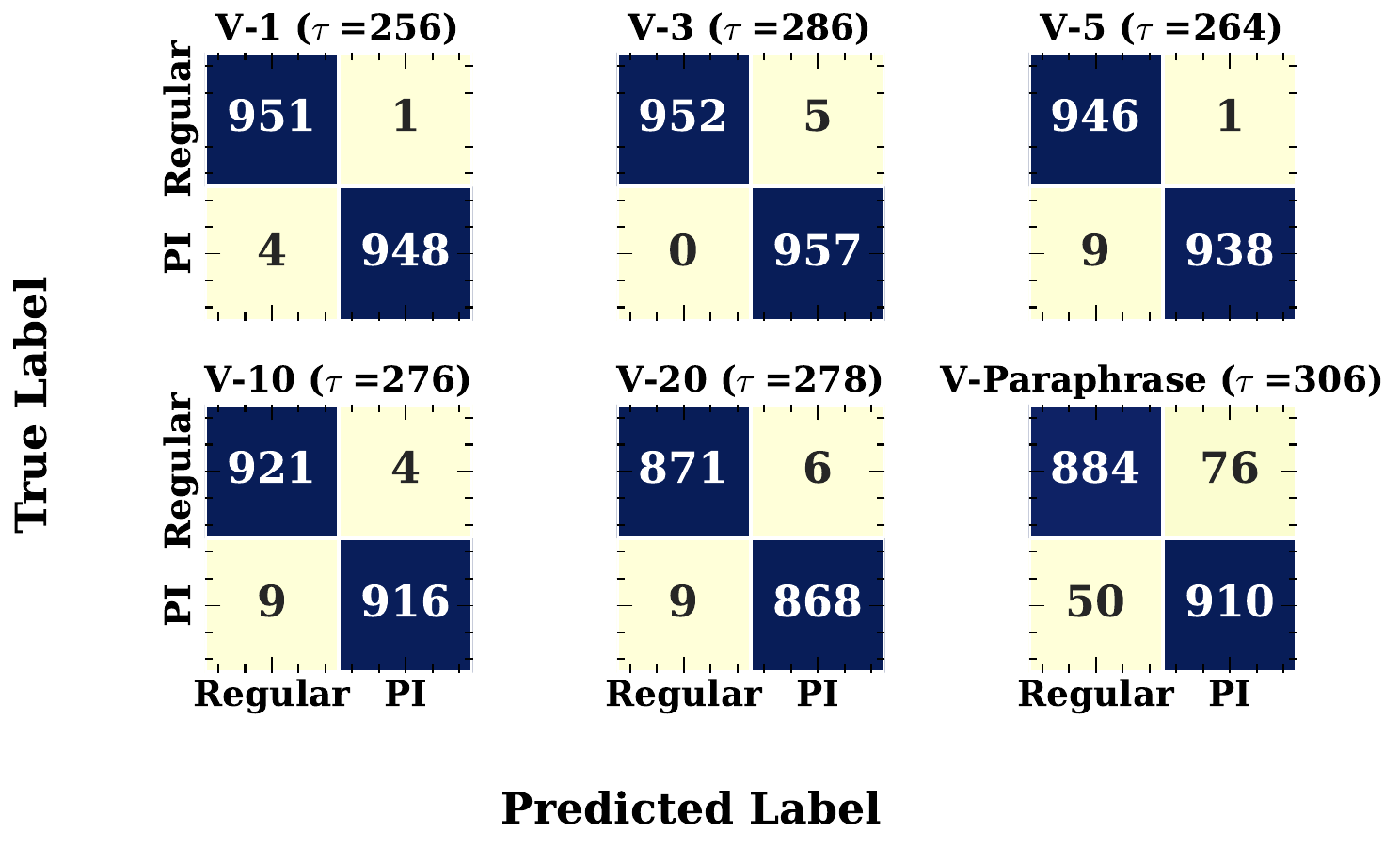}
        \caption{Confusion matrices at optimal thresholds.}
        \label{fig:binaryshield_cm}
    \end{subfigure}
    \caption{\underline{\textbf{\tool}} performance analysis across attack variants with $\alpha = 2.0$. (a) Precision-recall curves demonstrate consistent performance across variants. (b) Confusion matrices at optimal Hamming distance thresholds show TP, TN, FP, and FN counts, indicating strong detection capabilities even under complex paraphrasing attacks. Note that \textbf{PI} in the confusion matrices stands for \emph{Prompt Injection} and is used to distinguish between attack and benign prompts.}
    \label{fig:binaryshield_results}
    \vspace{-4mm}  
\end{figure}

\begin{figure}[t]
    \centering
    \begin{subfigure}[b]{0.45\textwidth}
        \includegraphics[width=\textwidth]{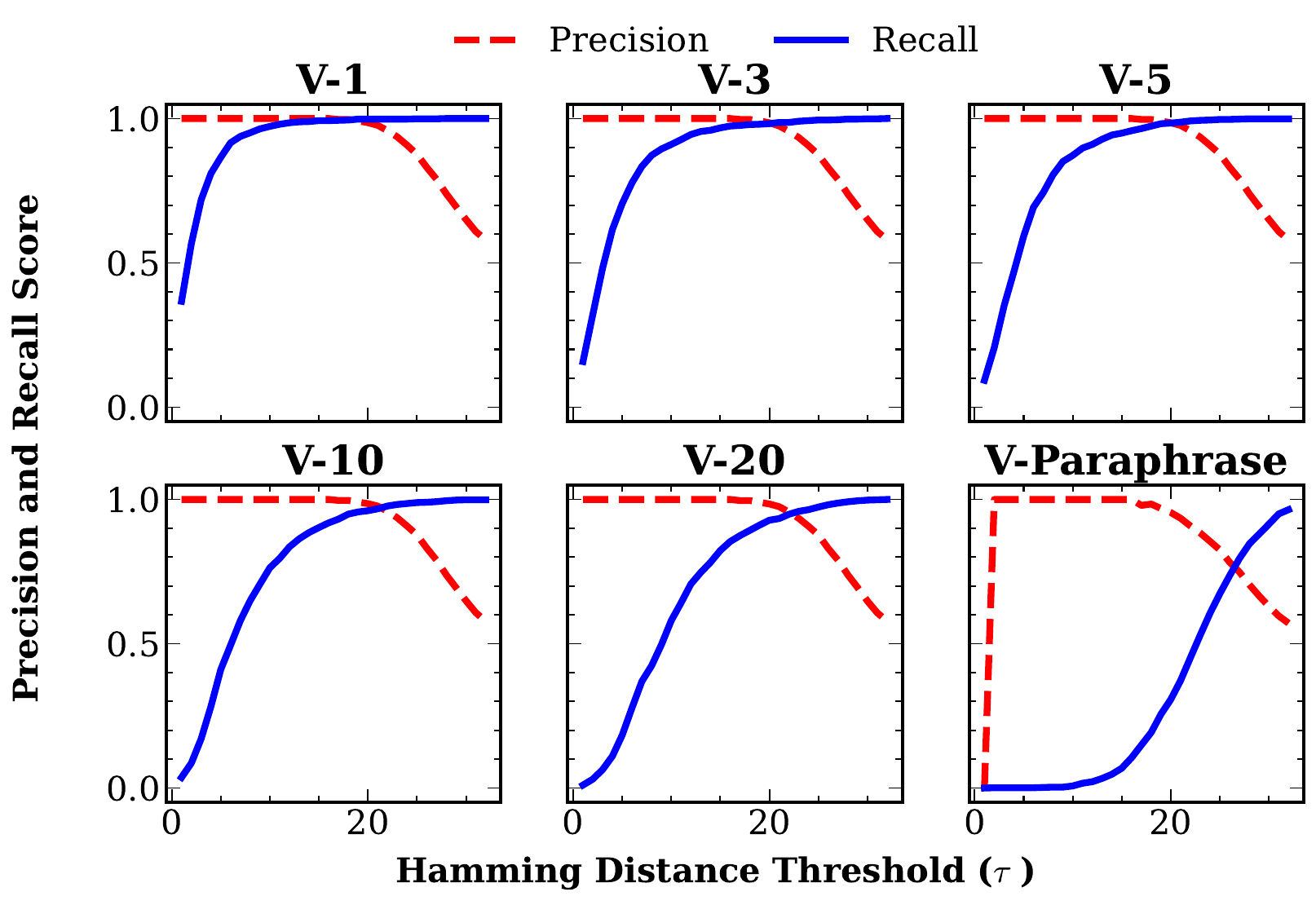}
        \caption{Precision-recall curves across attack variants.}
        \label{fig:simhash_pr}
    \end{subfigure}
    \hfill
    \begin{subfigure}[b]{0.45\textwidth}
            \centering
            \includegraphics[width=\textwidth]{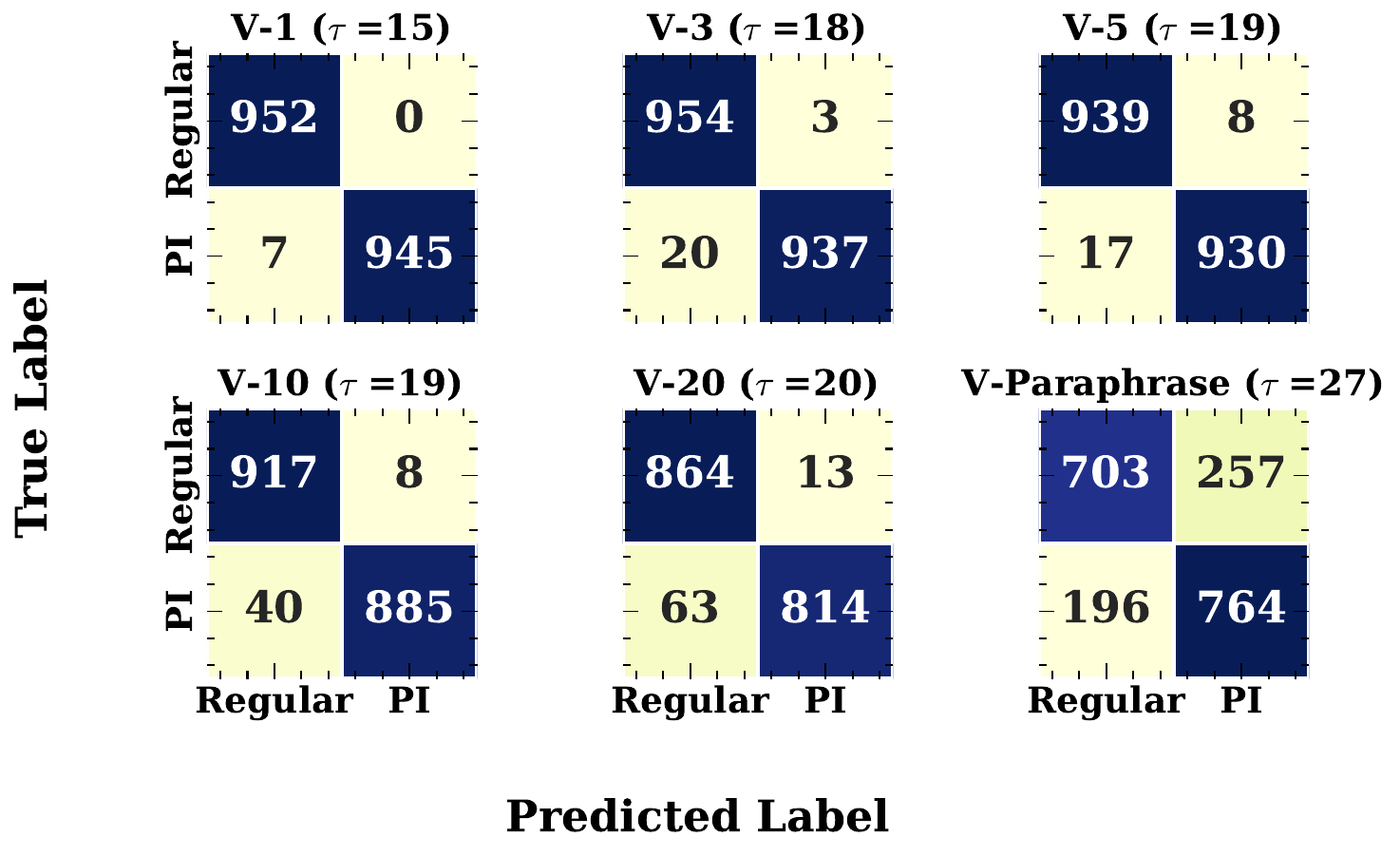}
        \caption{Confusion matrices at optimal thresholds.}
        \label{fig:simhash_cm}
    \end{subfigure}
    \caption{\underline{\textbf{SimHash}} baseline performance analysis across attack variants. Confusion matrices reveal growing FP and FN, particularly for paraphrasing scenarios.}
    \label{fig:simhash_results}
    \vspace{-4mm}  
\end{figure}

\subsection{\tool Comparison with Baseline}
\label{sec:comparison-baseline}

To provide a rigorous performance analysis, we systematically compare \tool against SimHash, a widely used locality-sensitive semantic hashing method~\cite{manku2007detecting}.
Both algorithms generate binary fingerprints and use Hamming distance for similarity measurement, making them directly comparable.
This baseline is chosen to evaluate not only detection performance but also the privacy-preserving capabilities essential for cross-boundary threat intelligence sharing as SimHash also generates privacy-preserving binary fingerprints.

For this \emph{evaluation}, we configure \tool with a differential privacy parameter of $\alpha = 2.0$, which provides meaningful privacy protection while maintaining strong detection performance.
Our comparative analysis encompasses six distinct attack variants: minimal single-word modifications (V-1), progressive word changes (V-5, V-20) and extensive modifications in paraphrasing scenarios (V-Paraphrase) that fundamentally restructure prompts while preserving malicious intent.
This progression allows us to assess how each method's performance degrades as attack sophistication increases during threat correlation.
Figure~\ref{fig:binaryshield_results} and Figure~\ref{fig:simhash_results} present comprehensive performance analysis through precision-recall curves and confusion matrices at optimal operating points.
\begin{figure}
    \centering
    \includegraphics[width=0.75\linewidth]{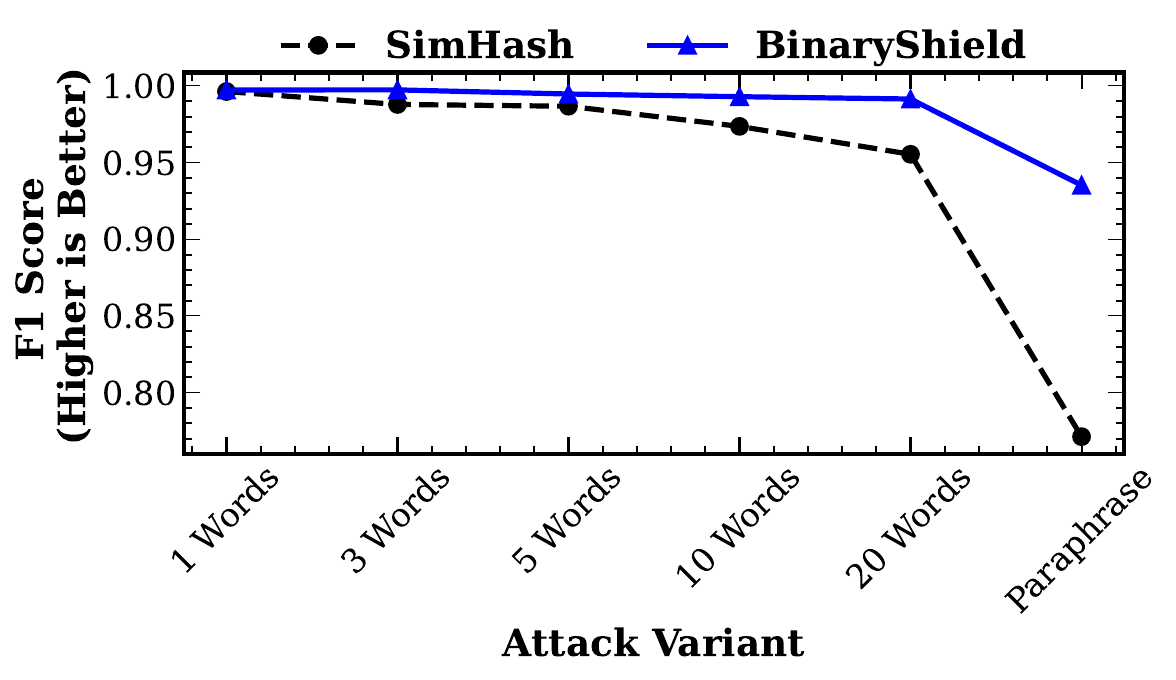}
    \caption{F1-score comparison between \tool and SimHash across attack variants. \tool outperforms SimHash, particularly in complex paraphrasing scenarios, demonstrating superior robustness to semantic modifications.}
    \label{fig:simhash_vs_binaryshield_f1}
    \vspace{-4mm}  
\end{figure}

For minimal attack modifications (V-1 to V-10), both algorithms demonstrate exceptional detection capabilities with near-perfect performance (Figure~\ref{fig:simhash_vs_binaryshield_f1}).
At optimal thresholds, both methods achieve optimal F1-scores (Figure~\ref{fig:simhash_vs_binaryshield_f1}), establishing a strong baseline for more complex evaluations.
For instance, in the V-10 scenario, SimHash produces 8 false positives and 40 false negatives at Hamming distance threshold ($\tau$) 19 (Figure~\ref{fig:simhash_results}), while \tool generates only 4 false positives at threshold 276 (Figure~\ref{fig:binaryshield_results}).

As attack complexity increases to extensive word modifications (V-20), performance differences begin to emerge.
\tool maintains robust performance with an F1-score of 0.99 (precision: 0.99, recall: 0.98), while SimHash experiences measurable degradation to an F1-score of 0.96 (precision: 0.98, recall: 0.93).
The 3.3\% gap starts to illustrate \tool's superior stability as lexical modifications increase.

The most challenging scenario involves comprehensive paraphrasing attacks (V-Paraphrase), where prompts undergo fundamental restructuring while preserving malicious intent.
Here, the performance differential becomes most pronounced: \tool achieves an F1-score of 0.94 (precision: 0.92, recall: 0.95), while SimHash degrades significantly to an F1-score of 0.77 (precision: 0.75, recall: 0.80). This 17\% gap in F1 score demonstrates \tool's fundamental advantage in threat intelligence that transcends surface-level linguistic variations.

Additionally, examining \emph{confusion matrices} at optimal thresholds, provides additional insights into error patterns in Figures ~\ref{fig:binaryshield_cm} and ~\ref{fig:simhash_cm}. \tool demonstrates consistent performance as compared to SimHash. SimHash shows increasing error variability as attack complexity increases, struggling particularly to distinguish between semantically similar malicious content and unrelated benign prompts.
The 17 percentage point F1-score gap in paraphrasing attacks, the most challenging and operationally relevant scenario, validates the effectiveness of \tool.
These performance improvements are achieved while simultaneously providing privacy guarantees essential for cross-service threat intelligence.

\begin{tcolorbox}[colback=white, colframe=black,  left=0pt, right=0pt, top=0pt, bottom=0pt]
    \textbf{Summary.}\ \tool consistently outperforms SimHash in threat correlation, especially under complex paraphrasing attacks, achieving an F1-score of 0.94 compared to SimHash's 0.77. These results highlight \tool's superior robustness with enabling privacy, making it highly effective for cross-service threat intelligence.
\end{tcolorbox}

\begin{figure}[t]
    \centering
    \includegraphics[width=\linewidth]{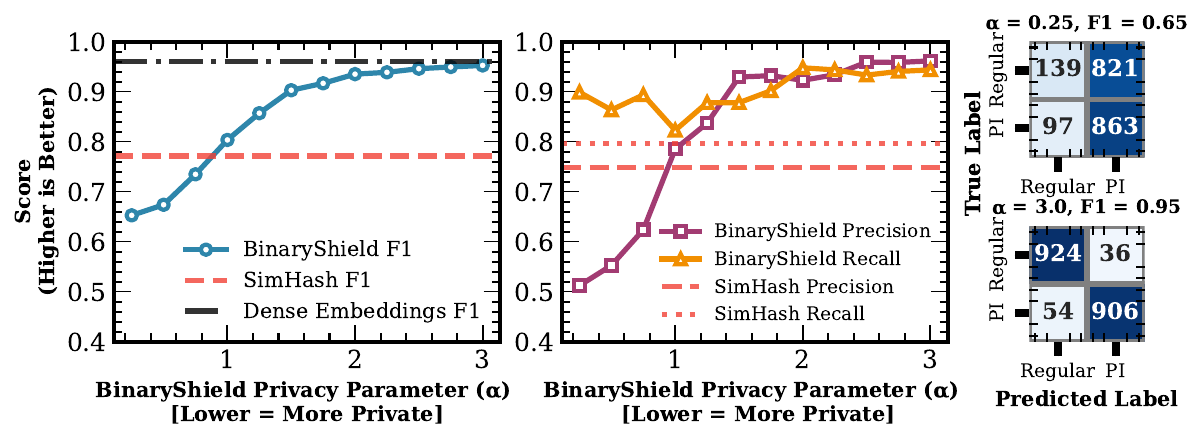}
    \caption{Privacy-utility trade-off for paraphrased prompt injection detection. (Left) \tool F1 rises smoothly with privacy budget $\alpha$, dominating SimHash (privacy-preserving baseline) across most of the spectrum and approaching non-private dense embeddings. (Center) Precision lags recall at high noise, then rapidly converges after $\alpha\!\approx\!1$ as FP collapse. Horizontal dashed lines: SimHash precision/recall. (Right) Confusion matrices at extreme settings: severe noise ($\alpha=0.25$) induces many FP; higher $\alpha$ ($3$) sharply improves precision.}
    \label{fig:binaryshield-privacy-utility-tradeoff-with-baselines}
    \vspace{-4mm}  
\end{figure}

\subsubsection{Empirical Analysis of Randomized Response Impact}
\label{sec:empirical-randomized-response}

\begin{figure}
    \centering
    \includegraphics[width=0.7\linewidth]{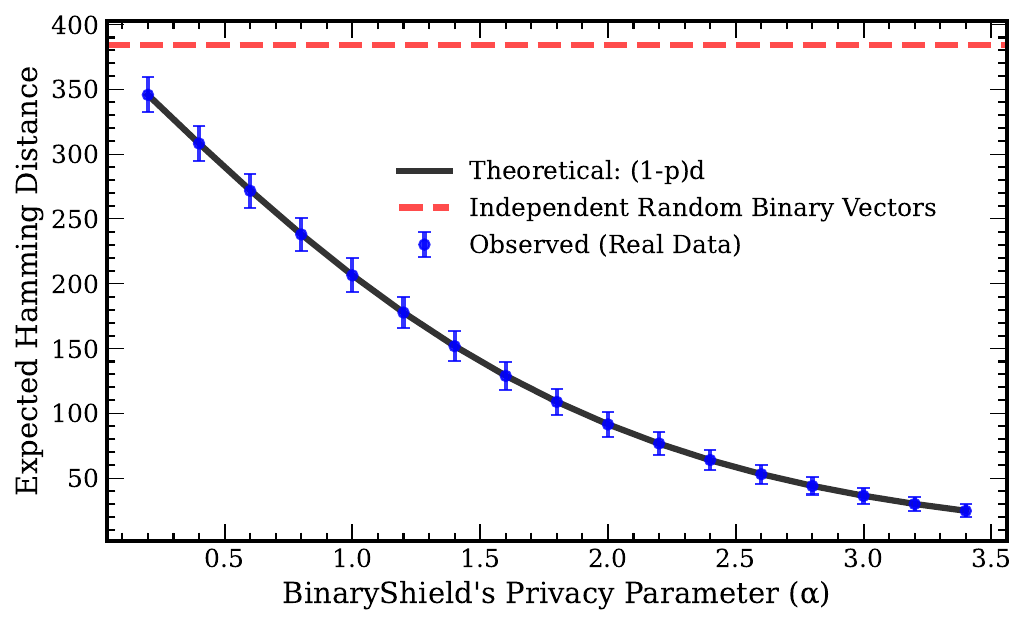}
    \caption{Calibration of randomized-response noise. Empirical mean Hamming distances (blue points, $n=500$ each) between original and privatized 768-bit fingerprints tightly follow the theoretical curve $(1-p)d$ (solid black) across privacy budgets $\alpha$.
        The dashed red line shows the mean distance between independent random binary vectors (384), illustrating how increasing $\alpha$ moves \tool smoothly from near-randomization to low-distortion states.}
    \label{fig:binaryshield-noise-calibration}
    \vspace{-20pt} 
\end{figure}

We quantify the impact of randomized response mechanism~\cite{warner1965randomized} noise on detection utility by sweeping the per‑bit privacy budget $\alpha\!\in\![0.25,3.0]$ (Algorithm~\ref{alg:binaryshield-pipeline}, Step 4) on the most challenging paraphrase attack variant.
For each $\alpha$, a bit in the 768‑dimensional binary embedding is kept with probability $p=\tfrac{e^{\alpha}}{e^{\alpha}+1}$ and flipped otherwise, yielding an expected self‑Hamming distortion $(1-p)d$, where $d$ is the fingerprint dimension (here, $d=768$).
At $\alpha=0.25$, $p=0.562$ and the mechanism flips $\approx0.438\times768\approx336$ bits (i.e., the whole binary vector is approximately completely randomized).
As shown in the top-right confusion matrix of Figure~\ref{fig:binaryshield-privacy-utility-tradeoff-with-baselines}, at this extreme privacy setting there are 821 false positives and 97 false negatives, resulting in an F1-score of only 0.65.
Such high noise levels provide maximal privacy but render the binary fingerprints nearly random (Figure~\ref{fig:binaryshield-noise-calibration}), offering little practical utility for threat detection.

Figure~\ref{fig:binaryshield-privacy-utility-tradeoff-with-baselines} shows that utility improves smoothly with larger $\alpha$ as noise shrinks.
At $\alpha=1.0$ precision increases to 0.79 with high recall of 0.82.
The inflection region $\alpha\!\in\![1.25,1.75]$ marks the transition where precision catches up to recall: by $\alpha=1.5$ \tool attains F1=0.90, already exceeding SimHash’s paraphrase F1 (0.77) by +13 points while providing privacy that is absent in dense cosine embeddings.
From $\alpha=2.5$ onward the F1 curve is close to the dense embeddings while still providing privacy as bit flipping probability is not zero (Figure~\ref{fig:binaryshield-noise-calibration}).
The confusion matrices at the privacy extremes (Figure~\ref{fig:binaryshield-privacy-utility-tradeoff-with-baselines}, right) illustrate this shift: moving from $\alpha=0.25$ to $3.0$ reduces false positives 22.8$\times$ (821$\rightarrow$36) and false negatives (97$\rightarrow$54), yielding an approximately 30‑point F1 gain (0.65$\rightarrow$0.95).
This predictable performance curve lets operators easily adjust $\alpha$ to meet privacy requirements, estimate detection accuracy, and securely share threat intelligence without revealing raw prompts or embeddings.

\noindent{\textbf{Randomized Response and Hamming Distance.}} To empirically validate \tool's privacy model, we conducted a controlled experiment measuring the actual Hamming distance introduced by the \tool's differential privacy.

To this end, for 500 prompts from our evaluation dataset, \tool generated 768-dimensional binary fingerprints and applied its differential privacy transformation by varying $\alpha$ values from 0.2 to 3.4.
For each configuration, we computed the Hamming distance between original and noise-perturbed fingerprints, comparing observed distances against the theoretical expectation $(1-p)d$.
As a baseline reference, we generated 1,000 pairs of independent random binary vectors to establish the average distance for completely uncorrelated binary vectors. The mean Hamming distance for these independent vectors is $384.44 \pm 13.65$, which serves as a baseline for comparison.
Figure~\ref{fig:binaryshield-noise-calibration} demonstrates that our theoretical model precisely predicts the observed self-Hamming distortion across all privacy levels confirming \tool's alignment with theory. At $\alpha = 0.2$, \tool's perturbed binary fingerprint reaches a Hamming distance of $346 \pm 13.41$, nearly matching the random baseline. This results in lower F1 scores at small $\alpha$ (Figure~\ref{fig:binaryshield-privacy-utility-tradeoff-with-baselines}).

As $\alpha$ increases, the Hamming distance decreases to $206.45 \pm 13.08 $ at $\alpha = 1.0$. This decay directly correlates with the smooth utility recovery in Figure~\ref{fig:binaryshield-privacy-utility-tradeoff-with-baselines}.
The tight correspondence between theoretical predictions and observed measurements validates that \tool's privacy mechanism operates exactly as designed, providing operators with precise control over the privacy-utility trade-off through a single, well-characterized parameter $\alpha$.
Another noticeable observation is that even at higher $\alpha$ values, as shown in Figure~\ref{fig:binaryshield-noise-calibration}, the Hamming distance remains nonzero. This indicates that \tool continues to provide privacy protection, since the nonzero Hamming distance ensures privacy is maintained, even as its utility approaches that of dense embeddings (Figure~\ref{fig:binaryshield-privacy-utility-tradeoff-with-baselines} and~\ref{fig:binaryshield-noise-calibration}).
In contrast, dense embeddings offer no privacy at all, so even a nonzero Hamming distance (i.e., some privacy) is fundamentally better than nothing, ensuring that sensitive information is not directly exposed.

\begin{tcolorbox}[colback=white, colframe=black,  left=0pt, right=0pt, top=0pt, bottom=0pt]
    \textbf{Summary.}\ \tool exhibits the fundamental differential privacy trade-off: low $\alpha$ (strong privacy) yields near-random fingerprints, whereas increasing $\alpha$ (meaningful privacy) produces a smooth, monotonic F1 rise that quickly approaches non-private baseline performance.
\end{tcolorbox}


\begin{figure*}
    \centering
    \includegraphics[width=\linewidth]{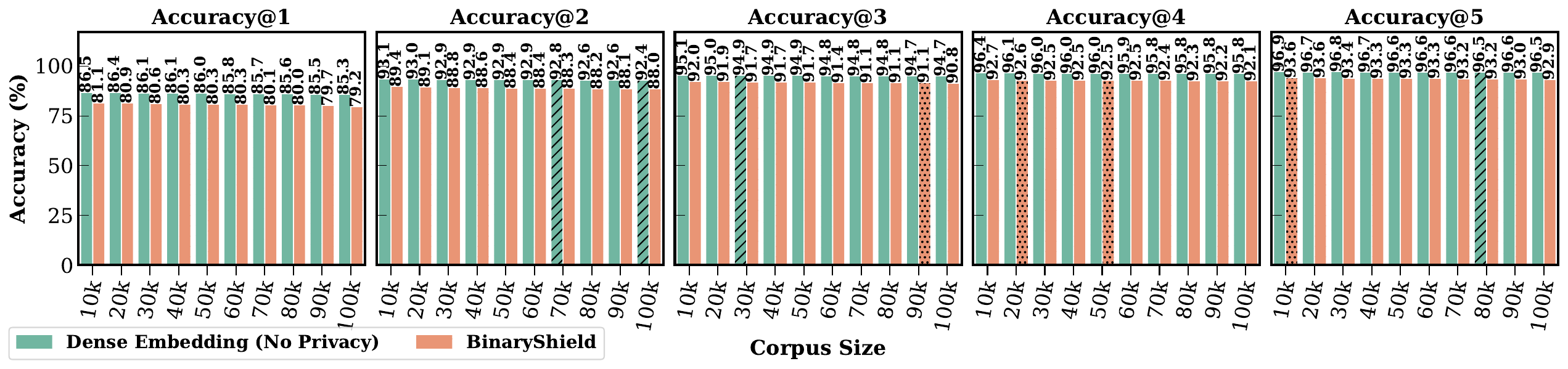}
    \caption{Scalability of \tool.
        Accuracy@k for \tool vs \emph{non-private} dense embeddings baseline across hybrid corpora (10K-100K). \tool retains 93\% of baseline Accuracy@1 while providing privacy.}
    \label{fig:hybrid-accuracy-scalability}
    \vspace{-4mm}  
\end{figure*}

\subsection{Real-World Deployment Analysis}
\label{sec:real-world-analysis}

While earlier sections validate \tool's superior performance as compared to the privacy-preserving baseline (SimHash), we now turn to evaluating \tool's effectiveness in realistic cross-service threat intelligence scenarios.
Specifically, it must be shown that privacy-preserving correlation remains effective when (i) malicious prompts are sparse relative to benign traffic and (ii) corpus size scales by an order of magnitude  (iii) privacy-utility trade-off is maintained, i.e., the accuracy gap between \tool and non-private dense embeddings remains bounded, and (iv) efficiency is maintained in terms of computational and storage overhead.

\subsubsection{Large-Scale Threat Intelligence Scalability}
\label{sec:large-scale-threat-intelligence-scalability}
To simulate realistic enterprise conditions, we query attack fingerprints against corpora where benign prompts vastly outnumber malicious ones, measuring Accuracy@k as successful detection when the true variant ranks within the top-k retrieved candidates. The goal is to quantify how \tool's privacy-preserving fingerprint correlation accuracy behaves as the total corpus grows, and characterize the accuracy gap relative to a non-private dense embedding baseline. In this experiment, \tool uses OpenAI's  \texttt{text-embedding-3-large} to generate embeddings. 

We study this setting using hybrid corpora that interleave a (fixed) malicious prompt injection (\texttt{V-Paraphrase}) with large volumes of real user interactions from WildChat~\cite{zhao2024wildchat}, thereby emulating an enterprise log in which only a small fraction of entries are attacks. For each target corpus size $C \in \{10\text{K},20\text{K},\dots,100\text{K}\}$, we construct a hybrid dataset by injecting a fixed malicious prompt set into $C - M$ benign WildChat prompts (where $M$ is the number of distinct attack prompts).
Each malicious prompt is issued as a query; success is recorded if its exact counterpart (or semantically identical variant, depending on $k$) appears within the top-$k$ candidates when ranked by Hamming distance over \tool's fingerprints, or cosine similarity for the dense baseline. We fix the local differential privacy parameter at $\alpha = 2$ (different $\alpha$ values are also evaluated in Section~\ref{sec:scalability-privacy-utility}).

Figure~\ref{fig:hybrid-accuracy-scalability} shows the results of threat intelligence scalability analysis, with the X-axis showing corpus sizes (10K to 100K) and the Y-axis representing retrieval accuracy.
\emph{Accuracy@k represents the percentage of queries for which the correct malicious prompt is returned in the first k-results.}
Accuracy@1 for the dense baseline declines modestly by only 1.2\% points across a 10$\times$ increase in corpus size (86.5\% $\rightarrow$ 85.3\%).
\tool{} shows a similarly shallow decline (81.1\% $\rightarrow$ 79.2\%; 1.9\% points). This indicates that neither binary quantization nor randomized response mechanism introduces scale-sensitive degradation.

Furthermore,~\Cref{fig:hybrid-accuracy-gap} (in the appendix) shows that the Accuracy@k \emph{gaps} between \tool and the dense baseline with X-axis showing corpus sizes (10K to 100K) and Y-axis showing the gap at different top-k accuracy.
The gap remains tightly bounded: it ranges from 5.31\% to 6.15\% (mean 5.66\% points) at $k=1$, indicating that \tool retains over 93\% of the dense baseline's accuracy even under tight local differential privacy constraints.
Crucially, the gap does \emph{not} significantly widen with corpus size.
Furthermore, increasing $k$ rapidly closes the gap, as shown in Figure~\ref{fig:hybrid-accuracy-scalability} and Figure~\ref{fig:hybrid-accuracy-gap}.
For instance at 100K entries, \tool achieves 79.2\% Accuracy@1, 90.8\% Accuracy@3, and 92.9\% Accuracy@5. On the same 100K corpus, the dense embeddings achieve 85.3\% Accuracy@1, 94.7\% Accuracy@3, and 96.5\% Accuracy@5.
The gap at $k=5$ is only 3.6\% points, indicating that \tool's with  privacy feature reaches close to the \emph{non-private} baseline of dense embeddings.
This confirms that the semantic signal preserved after (i) sign-only quantization and (ii) per-bit randomized response still remains sufficiently rich for practical threat correlation.
For every corpus size, \tool's incremental gains from $k=1$ to $k=2$ average roughly +8\% (Figure~\ref{fig:hybrid-accuracy-scalability}).
This reflects that most residual misses at $k=1$ are near-boundary semantic neighbors retrievable with minimal expansion. Subsequent marginal gains ($k>2$) diminish smoothly.
Figure~\ref{fig:hybrid-accuracy-gap} shows the gap curves monotonically compress as $k$ increases.
The gap is attributable to two irreversible transformations: (a) many-to-one sign mapping (loss of magnitude information) and (b) randomized response bit flips calibrated by $\alpha=2$.
That the gap does not drift upward with scale empirically supports that these transformations in \tool behave as \emph{scale-neutral perturbations} rather than compounding sources of semantic erosion.
In an operational setting where an analyst (or automated response engine) can examine the top 3--5 correlated candidates, \tool delivers privacy-preserving threat intelligence with accuracy comparable to a fully non-private dense embedding pipeline.
This constitutes strong utility under privacy constraints and validates \tool's suitability for multi-service deployment.

\begin{tcolorbox}[colback=white, colframe=black,  left=0pt, right=0pt, top=0pt, bottom=0pt]
    \textbf{Summary.} \tool retains over 93\% accuracy of dense embeddings (\emph{non-private baseline}) and remains scalable in practical deployment settings with randomized response mechanism, demonstrating robust privacy-preserving threat correlation at enterprise scale.
\end{tcolorbox}

\subsubsection{\tool's Privacy Parameter Performance at Scale}
\label{sec:scalability-privacy-utility}

\begin{figure}
    
    \centering
    \includegraphics[width=0.8\linewidth]{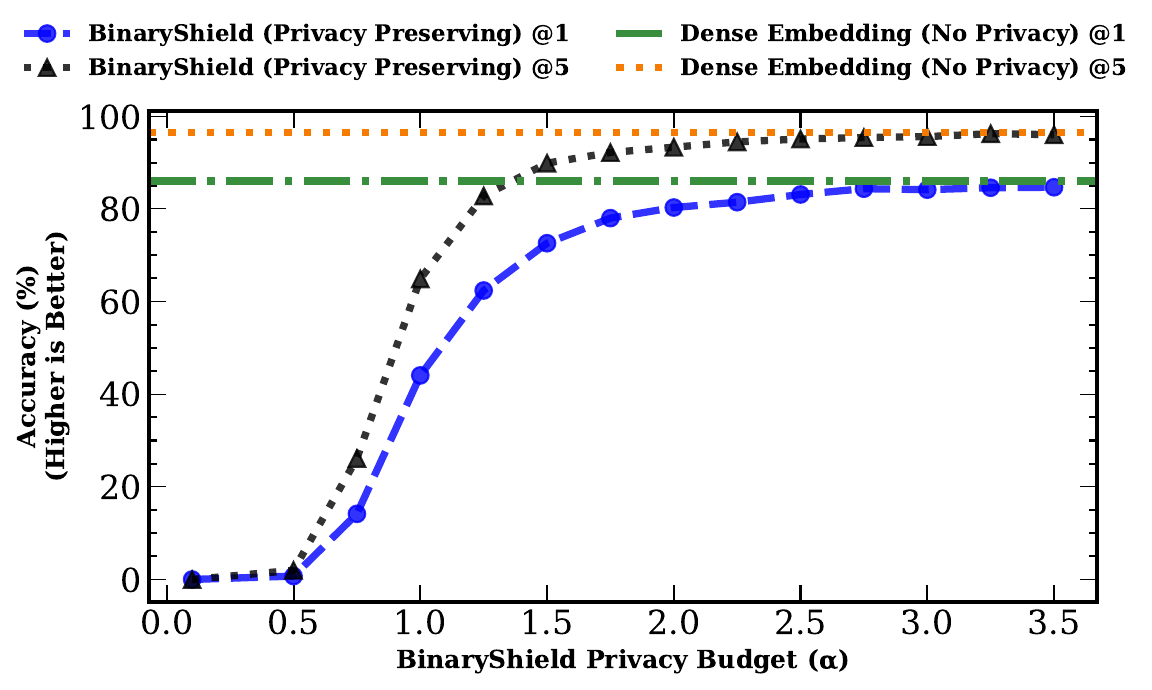}
    \caption{Privacy-utility trade-off for \tool across $\alpha$ values on 50K hybrid corpus.
        Accuracy@1 (blue line) and Accuracy@5 (black line) show phase transition behavior as privacy budget increases.
        }
    \label{fig:Scalability-Privacy--Utility-Trade-off-Characterization}
    \vspace{-4mm}  
\end{figure}

The relationship between privacy protection and threat detection utility at scale represents a fundamental tension in the cross-boundary threat intelligence system.
Similar to Section~\ref{sec:empirical-randomized-response}, to characterize this trade-off precisely, we evaluate \tool's performance across a spectrum of privacy budgets ($\alpha$) on a fixed 50K-entry hybrid corpus containing both malicious prompts and benign WildChat interactions (Section~\ref{sec:large-scale-threat-intelligence-scalability}).
The privacy parameter $\alpha$ directly controls the randomized response mechanism through the bit preservation probability $p = e^{\alpha}/(e^{\alpha} + 1)$, with the corresponding bit flip probability being $1 - p$.
Smaller $\alpha$ values provide stronger privacy guarantees by increasing the probability of bit flips in the binary fingerprint, while larger values preserve more semantic information at the cost of reduced privacy protection.

Our experimental results, visualized in Figure~\ref{fig:Scalability-Privacy--Utility-Trade-off-Characterization}, reveal a sharp phase transition in detection accuracy as the privacy budget increases from highly restrictive to meaningful protective settings. The X-axis represents the privacy parameter $\alpha$, while the Y-axis shows correlation accuracy (Accuracy@1 and Accuracy@5) for both \tool and the \emph{non-private} dense embedding baseline.

At extreme privacy levels ($\alpha = 0.1$), the bit flip probability reaches 0.475, meaning nearly half of all bits are randomly inverted, effectively reducing fingerprints to near-random noise.
Consequently, Accuracy@1 drops to 0\%, rendering threat correlation impossible.
This represents the theoretical limit where maximal privacy completely eliminates utility. As we relax the privacy constraint the bit flip probability decreases.
With $\alpha = 0.75$ (bit flip probability 0.32), Accuracy@1 rises to 14.17\% and Accuracy@5 to 26.15\%, marking the onset of effective threat detection.
At $\alpha = 1.0$  \tool to reach 44.06\% Accuracy@1 and 64.90\% Accuracy@5.
While this is roughly half the non-private baseline (86.04\% and 96.56\%), it shows that meaningful threat correlation is feasible even under strong privacy constraints.
Between $\alpha = 1.0$ and $\alpha = 2.0$, accuracy improves smoothly with meaningful privacy and utility tradefoff.
At $\alpha = 1.25$, Accuracy@1 reaches 62.40\% and Accuracy@5 82.81\%.
At $\alpha = 1.5$, Accuracy@1 is 72.60\% and Accuracy@5 is 89.90\%, retaining 84.4\% of baseline Accuracy@1.

The results show a smooth transition in the privacy-utility trade-off, especially between $\alpha = 0.75$ and $\alpha = 2.0$.
Since organizations have already strict security policies, for internal sharing, relaxing privacy to $\alpha = 2.0$--$2.5$ yields near-optimal utility with \emph{realistic} meaningful privacy protection. Overall, \tool potentially can operate in a wide range of privacy-utility trade-offs.

\begin{tcolorbox}[colback=white, colframe=black,  left=0pt, right=0pt, top=0pt, bottom=0pt]
    \textbf{Summary.} \tool empowers organizations to balance privacy and utility as needed, consistently delivering strong accuracy and realistic privacy protection. The organizations can flexibly tune the privacy-utility trade-off based on their operational and regulatory needs.
\end{tcolorbox}

\subsubsection{\tool Computational Efficiency Analysis}
\label{sec:computational-storage-efficiency}

The operational viability of any threat intelligence system fundamentally depends on its ability to process vast quantities of data with minimal latency and storage overhead.
Historical lessons from traditional cybersecurity infrastructure underscore this requirement.
One of the major reasons that signature-based ecosystems (antivirus, IDS) scaled is because compact hashes (e.g., MD5,  fuzzy-hash families).
Similarly, modern LLM services face an even more daunting challenge, processing billions of queries daily across distributed infrastructure.
For \tool to enable practical cross-boundary threat intelligence, it must achieve orders-of-magnitude improvements in both computational speed and storage efficiency compared to \emph{non-private} dense embeddings.
\emph{Every extra byte per fingerprint multiplies into material storage cost, and every millisecond of per-fingerprint scan latency stretches incident response time}.

\begin{figure}[t]
     \centering
        \includegraphics[width=0.7\linewidth]{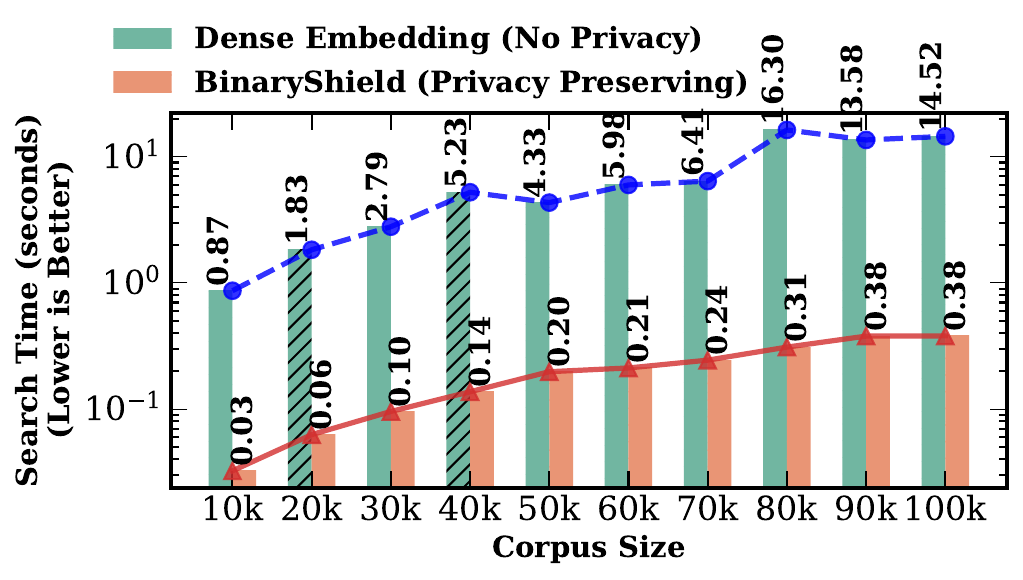}
        \caption{Search time comparison between non-private dense embeddings and \tool across corpus sizes. Note the logarithmic scale on the Y-axis. \tool maintains sub-second performance, demonstrating its suitability for real-time threat correlation at enterprise scale.}
        \label{fig:search-time-comparison}
        \vspace{-4mm}  
\end{figure}

Figure~\ref{fig:search-time-comparison} reveals \emph{computational advantages} of \tool's fingerprints over dense embeddings. At a corpus size of 10K entries, dense embeddings require 0.87 seconds for similarity search of 968 prompt injections, while \tool completes the same operation in just 0.032 seconds, a 26.7x speedup. This performance gap widens dramatically as corpus size increases. At 100K entries, dense embeddings demand 14.52 seconds, rendering real-time threat correlation infeasible for high-volume services. In contrast, \tool maintains sub-second performance at 0.38 seconds, achieving a 38.1x speedup. The computational efficiency stems from the fundamental difference in similarity computation. Dense embeddings require floating-point dot products or cosine similarity calculations, operations that scale poorly even with optimized linear algebra libraries. \tool's binary fingerprints enable Hamming distance computation through simple XOR operations. We discuss \tool's storage utilization in the appendix. 

\emph{The practical implications of these efficiency gains extend beyond raw performance metrics (Section ~\ref{sec:comparison-baseline} and ~\ref{sec:large-scale-threat-intelligence-scalability}).}
Consider an organization processing 100 million queries daily across ten services. With dense embeddings, maintaining a rolling 30-day threat intelligence window would require approximately 7.1 TB of storage per boundary and dedicated \emph{GPU clusters} for similarity search making it near to impractical choice. \tool reduces this to 111 GB per boundary, fitting comfortably in memory on commodity servers while enabling CPU-based similarity search. This efficiency democratizes threat intelligence capabilities, allowing even resource-constrained services to participate in collaborative defense without significant infrastructure investment.

Furthermore, the computational efficiency enables new operational capabilities. 
Security teams can now perform retrospective threat hunting across months of historical data in minutes rather than hours.
The reduced storage footprint also facilitates comprehensive threat intelligence archival, enabling long-term trend analysis and attribution of persistent attack campaigns that evolve over extended periods.

\begin{tcolorbox}[colback=white, colframe=black,  left=0pt, right=0pt, top=0pt, bottom=0pt]
    \textbf{Summary.} \tool's computational and storage efficiency fundamentally transforms cross-boundary threat intelligence economics. 
    By reducing search latency up to 38x and storage requirements while maintaining 93\% of non-private baseline top-1 accuracy with privacy, \tool enables enterprise-scale threat correlation. 
    This positions \tool as a practical foundation for collaborative threat intelligence systems against prompt injection attacks. 
\end{tcolorbox}

\section{Related Work and Discussion}

\noindent{\textbf{Attacks.}}
Several works have focused on systematically characterizing prompt injection attacks.
For instance, \cite{greshake2023not} demonstrate how hidden instructions can be inserted into external content to trigger harmful model behaviors.
Other works~\cite{wei2023jailbroken, zou2023universaltransferableadversarialattacks, zhang2024effective, deng2023masterkey, huang2024catastrophic, liu2024autodan, shen2024doanything, deng2024multilingual, chao2025jailbreaking} further contribute to our understanding by exploring a variety of prompt injection scenarios.
In addition,~\cite{Yang2024SOS} propose \texttt{SOS}, a training-time \say{soft prompt} attack that implants backdoors into open-source LLMs. The LLM behaves normally, until a trigger token activates malicious behaviors like jailbreaks, prompt stealing, or output manipulation.
In contrast,~\cite{russinovich2025greatwritearticlethat} introduce \texttt{Crescendo}, an inference-time jailbreak attack that unfolds over multiple dialogue turns.
Supporting this observation,~\cite{bullwinkel2025representation} performs an in-depth analysis of such multi-turn jailbreak attacks and show why defenses designed for single-turn interactions, like circuit breakers \cite{zou2024improving}, may be insufficient.
By examining internal model representations,~\cite{bullwinkel2025representation} shows that multi-turn prompts can gradually reframe harmful outputs as safe, effectively evading safeguards designed for single-turn interactions.
\texttt{BIPIA}~\cite{yi2025benchmarking} systematically assesses the risks posed by malicious instructions embedded in third-party content and introduces two defense paradigms: one based on prompt engineering and in-context strategies in a black-box setting, and another using adversarial training with special tokens in a white-box setting.
Finally,~\cite{abdelnabi2025llmail} introduces a realistic testbed for indirect prompt injection attacks targeting an LLM-based email assistant.
Collectively, these studies~\cite{greshake2023not, wei2023jailbroken, zou2023universaltransferableadversarialattacks, zhang2024effective, deng2023masterkey, huang2024catastrophic, liu2024autodan, shen2024doanything, deng2024multilingual, chao2025jailbreaking, Yang2024SOS, russinovich2025greatwritearticlethat, bullwinkel2025representation, yi2025benchmarking, abdelnabi2025llmail} provide a comprehensive overview of the diverse strategies employed to exploit prompt injection attacks, underscoring the ongoing need for robust security measures in LLM-integrated applications and services.

\noindent{\textbf{Attack Defenses.}}
Recent work proposes a variety of defense strategies against prompt injection attacks. These approaches differ in their methods and operating settings, ranging from inference-time countermeasures to white-box techniques and architectural safeguards.
\texttt{RigorLLM}~\cite{yuan2024rigorllm} framework enhances content moderation. Hines et al.~\cite{Hines2024Defending} build on the work of ~\cite{yi2025benchmarking}
to address indirect prompt injection attacks by proposing a defense strategy based on \emph{spotlighting}.
\texttt{CaMeL}~\cite{Debenedetti2025defeating} is a defense mechanism that explicitly detects and prevents prompt injection attacks by designing a secure execution layer around LLM operations.
The work sits at the intersection of generative AI and software security, bringing established principles from the latter to protect emerging systems.
SafeEar~\cite{SafeEar} proposes a method for detecting deepfake audio without accessing the actual speech content.
\texttt{TaskTracker}~\cite{abdelnabi2025get} is a white-box defense,  that centers on tracking the model’s internal activations before and after processing external text. 
\texttt{DataSentinel}~\cite{liu2025datasentinel} detects prompt injection attacks by fine-tuning a language model with a game-theoretic mini-max optimization. It uses a detection instruction embedding a secret key to verify the integrity of incoming data.
Beurer-Kellner et al.~\cite{beurerkellner2025design} propose a six architectural design patterns (action-selector, plan-then-execute, map-reduce, dual LLM, code-then-execute, and context-minimization) that secure LLM agents against attacks. 
Overall, most existing defenses are probabilistic and can be bypassed~\cite{Debenedetti2025defeating, costa2025securingaiagentsinformationflow}. Retrospective systems are needed to identify attackers, yet cross-service threat intelligence for large-scale prompt injection attacks correlation remains unexplored. To address this gap, techniques like SimHash can be leveraged. However, SimHash captures only syntactic similarity and fails to capture semantic similarity (Figures~\ref{fig:binaryshield_results},~\ref{fig:simhash_results}, and~\ref{fig:simhash_vs_binaryshield_f1}). 
\emph{\tool is the first to focus on this gap, proposing a concrete system for cross-service threat intelligence and paving the way for future research in this domain.}

\subsection{Discussion}  
\label{sec:discussion-and-threats-to-validity}

Our study has limitations similar to those in any research. 
Synthetic paraphrasing and word substitution may yield varying results with different LLMs and prompts. 
The effectiveness of the \tool's PII redaction module depends on the accuracy of detection methods. 
Residual PII may remain in prompts.
Our results rely on specific embedding models, ModernBert (open-source)~\cite{warner-etal-2025-smarter, gill2025advancingsemanticcachingllms} and OpenAI text-embedding-3-large (proprietary). 
Thus, performance may vary with other models. 
We determined optimal Hamming distance thresholds through search on our datasets. 
However, these thresholds may not generalize to other attack distributions or organizational contexts.
All experiments contain English prompts. 
Future work should explore cross-lingual attack detection and domain-specific technical prompts. 
We depend on computational non-invertibility and differential privacy guarantees. 
The evaluation does not assess resistance to reconstruction attacks, which represent an orthogonal concern to current focus of the paper.

\noindent{\textbf{Differential Privacy.}} Noise can be added to the fingerprint through various differential privacy mechanisms~\cite{warner1965randomized, Wang2017Locally, bhowmick2018protection, rappor, feyisetan2021private, vadhan2021concurrent, carvalho2021brr, bollegala2023neighbourhood, xie2024differentially, meehan2022sentence, asi2022optimal, du2023sanitizing, bittau2017prochlo, fanti2016building, hsu2014differential, near_abuah_2021, fioretto2024differential, dwork2014algorithmic, pmlr-v80-balle18a, NIPS2017_253614bb, hu2024sok}. 
\tool utilizes foundational work of Randomized Response mechanism~\cite{warner1965randomized}. 
Assessing the impact of other DP mechanisms and global differential privacy on the overall privacy guarantees is an important direction for future research. 
We did not formulate the privacy utility trade off as an optimization problem. 
Readers may consult existing literature on this topic~\cite{near_abuah_2021, Wang2017Locally, dwork2014algorithmic, fioretto2024differential}. 
However, in our experiments, we empirically evaluated conditions ranging from extreme privacy, where accuracy approaches zero, to moderate and weak privacy.
Each organization has specific privacy policies, and we do not recommend a particular differential privacy budget. 
Organizations should work with their privacy teams to select an appropriate budget based on their guidelines.  

\section{Conclusion}
\label{sec:conclusion}

Regulatory boundaries prevent an enterprise's multiple LLM services from sharing prompt injection threat intelligence. This isolation means an attack blocked by one service can still succeed against another, as each service has its own defense mechanism. 
This creates a fundamental gap in the organization's collective defense capabilities.

\tool addresses this critical gap by introducing the first privacy-preserving fingerprinting mechanism specifically designed for prompt injection threat intelligence. Through a carefully orchestrated pipeline combining PII redaction, semantic embedding, binary quantization, and randomized response, \tool generates privacy-preserving fingerprints that can be safely shared across compliance boundaries while preserving the semantic characteristics necessary for effective threat correlation. 

The implications of this work extend beyond immediate operational benefits. As LLM-based systems become critical infrastructure across finance, healthcare, and government services, the attack surface for prompt injection continues to expand exponentially. The emergence of autonomous agents and MCP servers further amplifies these risks, potentially enabling attacks that cascade from text manipulation to arbitrary code execution and system compromise. \tool establishes the foundational system for collaborative defense against these evolving threats, creating a pathway toward industry-wide threat intelligence feeds analogous to existing malware signature ecosystems.

\section{LLM Usage Considerations}

We utilized GPT-4o to systematically generate the attack variants (word-flipped and paraphrased) for our evaluation. This approach was necessary because no public dataset exists for cross-service threat intelligence analysis and enterprise data access is restricted. To address this gap, we used GPT-4o to create a partially synthetic dataset from original prompt injection attacks. The generation was a one-time, offline process using efficient prompts (see Appendix \Cref{prompt:words-flipped} and \Cref{prompt:paraphrase}) to minimize the computational footprint. While using a closed-source model affects exact reproducibility, our generation process is transparent, and the resulting dataset characteristics are replicable with other powerful LLMs. We also briefly discuss this limitation in Section~\ref{sec:discussion-and-threats-to-validity}.

We also used LLMs to correct grammar mistakes, similar to a feature like Grammarly. The whole content in this study is written entirely by the authors, who take full responsibility for the paper.

\section*{Acknowledgment}
We sincerely thank the SaTML reviewers for their valuable feedback. 
We are also deeply grateful to Yonatan Zunger, Angela Argentati, Dharmin Shah, Sukirna Roy, Habiba Mohamed, Anandan Sundar, Leah Zulas, and the entire Microsoft AI Safety and Security organization for their support.

\bibliographystyle{IEEEtran}
\bibliography{main}

\appendix

\appendices

\subsection{Threat Model}
 An organization offers several siloed LLM services, each compartmentalized with distinct stacks and compliance boundaries. An adversary interacts with these services via standard prompt interfaces (Figure~\ref{fig:system-architecture}). The adversary has no access to internal artifacts (such as logs, fingerprints, or defense details) and cannot tamper with compliance boundaries. The adversary sends prompt-injection attack variants to exploit vulnerabilities across these services. In \tool, if a prompt injection attack is detected in one of these services, the goal is to share this threat intelligence with other peer services while preserving privacy. \tool shares a privacy-preserving fingerprint of the detected attack, which peer services can correlate in a privacy-preserving manner, thereby improving organization-wide visibility of attacks.

\subsection{Motivating Example}
The motivation in context of compliance boundaries is discussed in earlier sections (Section~\ref{sec:introduction} and Section~\ref{sec:design}). Here is another motivating example to demonstrate \tool effectiveness. 
Consider a cloud provider offering generative AI services to multiple independent enterprise customers, hereafter referred to as Company A and Company B. Note in this scenario companies are customers of the cloud provider. Due to strict regulatory requirements and contractual obligations, these two customers operate in completely separate compliance boundaries. Their data cannot be shared, and their logging systems are entirely siloed.
 
Now suppose a malicious actor launches prompt injection attacks against both companies' AI services. Because the data between Company A and Company B cannot be correlated or shared directly, each company's security team operates in isolation. They may detect anomalies independently, but they have no way to identify that the same attacker is targeting both services, nor can they assess the full scale and sophistication of the attack campaign.
 
This is where \tool provides value. By generating privacy-preserving fingerprints of detected malicious prompts, \tool enables the cloud provider to correlate threat intelligence across these two isolated compliance boundaries without exposing the original prompts or violating privacy regulations.

 \begin{table}[h]
\centering
\caption{Performance comparison on the WildJailbreak dataset.}
\label{tab:wildjailbreak}
\begin{tabular}{lccc}
\hline
Metric & Dense (No Privacy) & \tool & Gap \\
\hline
Accuracy@1 & 99.25\% & 93.82\% & 5.43\% \\
Accuracy@2 & 99.70\% & 97.54\% & 2.16\% \\
Accuracy@3 & 99.70\% & 98.54\% & 1.16\% \\
Accuracy@5 & 99.90\% & 99.3\% & 0.6\% \\
\hline
\end{tabular}
\end{table}

\begin{figure}
    \centering
    \includegraphics[width=\linewidth]{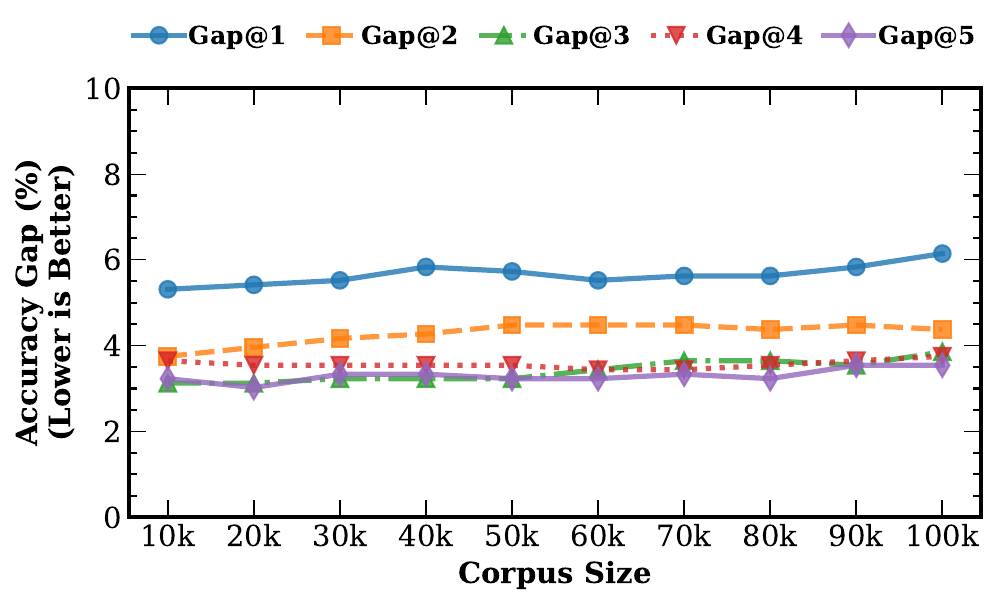}
    \caption{Accuracy@k gap (Dense Embeddings - \tool) across corpus sizes.  Rapid narrowing for higher k demonstrate that binary quantization plus local DP introduce a constant, not compounding, utility cost.}
    \label{fig:hybrid-accuracy-gap}
\end{figure}

\begin{figure}[t]
        \centering
        \includegraphics[width=\linewidth]{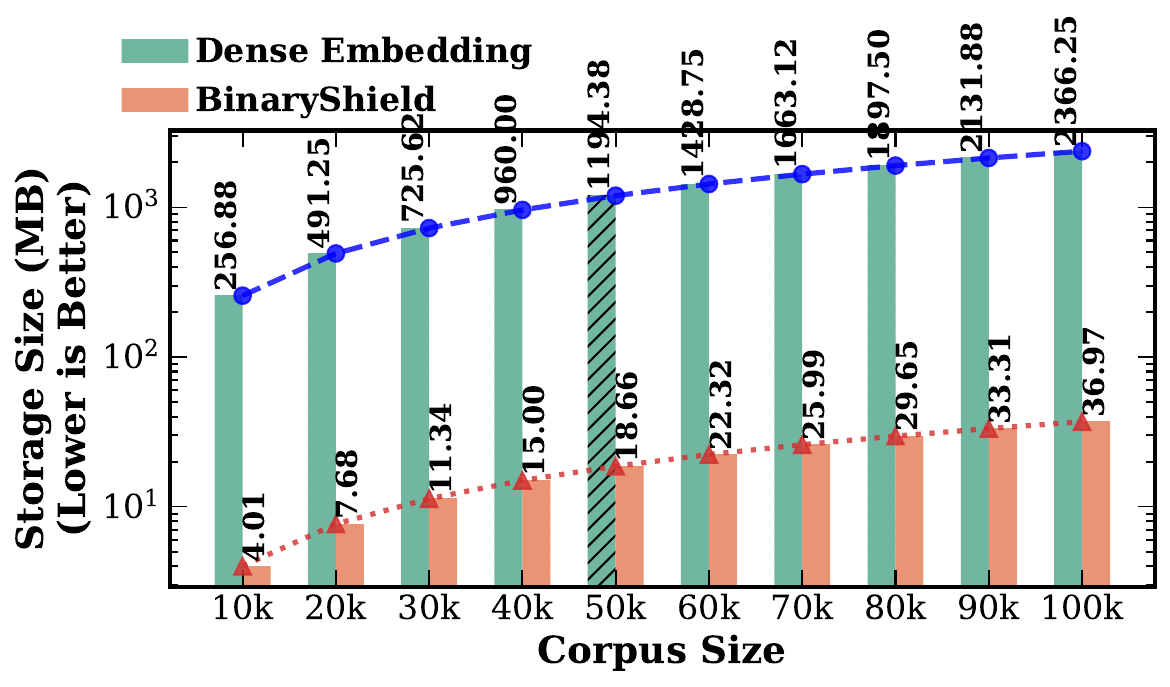}
        \caption{Storage size comparison demonstrating \tool's 64x reduction compared to dense embeddings. Dense embeddings scale from 256.88 MB at 10K entries to 2,366.25 MB at 100K entries, while \tool requires only 4.01 MB to 36.97 MB respectively, enabling cost-effective long-term threat intelligence retention.}
        \label{fig:storage-size-comparison}
    
\end{figure}

\begin{lstlisting}[style=customprompt, basicstyle=\ttfamily\scriptsize,caption={Template for generating adversarial prompt variants via lexical substitution.}, label={prompt:words-flipped}]
Understand this adversarial prompt: <original prompt>. Now generate synonyms for the following words: <words to change>. Return the JSON with the field `words_changes_to` containing the synonyms in order.
\end{lstlisting}

\begin{lstlisting}[style=customprompt, basicstyle=\ttfamily\scriptsize,caption={Template for generating adversarial prompt variants via full-sentence paraphrasing.}, label={prompt:paraphrase}]
Paraphrase the following adversarial prompt while maintaining the attack intent and purpose. Original prompt: <original prompt>. Return the JSON with the field `generated_response`.
\end{lstlisting}

\subsection{Synthetic Variant Attack Potency and Evaluation on WildJailbreak Dataset}
We conduct an evaluation to test the attack potency of our synthetic variants, specifically examining whether they maintain attack potency comparable to original attacks. We conduct a study comparing 500 original attack prompts~\cite{shen2024doanything} and their synthetic variants (V\_Paraphrase) using GPT-4.1-nano combined with Azure's Prompt Shields for attack detection in an enterprise setting. Original attacks achieve an 80.4\%  detection rate, while synthetic variants achieve a 78.0\%. These near-identical detection rates demonstrate that the synthetic variants preserve adversarial characteristics comparable to the original attacks.

Several jailbreak datasets exist~\cite{WildJailbreak, shen2024doanything, luo2024jailbreakv, JailbreaksOverTime}. We evaluate \tool on~\cite{shen2024doanything} in Section~III and on WildJailbreak~\cite{WildJailbreak} here similar to Section~III-C. Table~\ref{tab:wildjailbreak} confirms \tool's robustness on WildJailbreak. It achieves 93.82\% Accuracy@1 on a 100K corpus of benign and adversarial prompts, retaining 94.5\% of the non-private baseline, consistent with our earlier results (Figures~\ref{fig:hybrid-accuracy-scalability} \&~\ref{fig:hybrid-accuracy-gap}).

\subsection{Storage Efficiency}
The storage requirements present an equally compelling case for \tool's architecture, as illustrated in Figure~\ref{fig:storage-size-comparison}. Dense embeddings consume 256.88 MB for 10K entries, growing linearly to 2,366.25 MB at 100K entries. This represents a substantial infrastructure burden when organizations must maintain threat intelligence databases spanning billions of historical queries across multiple compliance boundaries. \tool reduces storage requirements by a factor of 64, requiring only 4.01 MB for 10K entries and 36.97 MB for 100K entries. Our reported storage figures use 64-bit floating-point precision. Storage requirements would be halved with the more commonly used 32-bit precision.
This 64x reduction directly translates from the binary quantization process, where each embedding dimension is compressed from a floating-point value to a single bit, followed by the addition of  noise through randomized response, which preserves the binary nature.

\end{document}